\documentclass[usenatbib]{mn2e}
\usepackage{amssymb, amsmath}
\usepackage{epsfig}
\usepackage{graphicx}
\usepackage{color}
\usepackage{epstopdf}
\definecolor{AliceBlue}{rgb}{0.94,0.97,1.00}
\definecolor{AntiqueWhite1}{rgb}{1.00,0.94,0.86}
\definecolor{AntiqueWhite2}{rgb}{0.93,0.87,0.80}
\definecolor{AntiqueWhite3}{rgb}{0.80,0.75,0.69}
\definecolor{AntiqueWhite4}{rgb}{0.55,0.51,0.47}
\definecolor{AntiqueWhite}{rgb}{0.98,0.92,0.84}
\definecolor{BlanchedAlmond}{rgb}{1.00,0.92,0.80}
\definecolor{BlueViolet}{rgb}{0.54,0.17,0.89}
\definecolor{CadetBlue1}{rgb}{0.60,0.96,1.00}
\definecolor{CadetBlue2}{rgb}{0.56,0.90,0.93}
\definecolor{CadetBlue3}{rgb}{0.48,0.77,0.80}
\definecolor{CadetBlue4}{rgb}{0.33,0.53,0.55}
\definecolor{CadetBlue}{rgb}{0.37,0.62,0.63}
\definecolor{CornflowerBlue}{rgb}{0.39,0.58,0.93}
\definecolor{DarkBlue}{rgb}{0.00,0.00,0.55}
\definecolor{DarkCyan}{rgb}{0.00,0.55,0.55}
\definecolor{DarkGoldenrod1}{rgb}{1.00,0.73,0.06}
\definecolor{DarkGoldenrod2}{rgb}{0.93,0.68,0.05}
\definecolor{DarkGoldenrod3}{rgb}{0.80,0.58,0.05}
\definecolor{DarkGoldenrod4}{rgb}{0.55,0.40,0.03}
\definecolor{DarkGoldenrod}{rgb}{0.72,0.53,0.04}
\definecolor{DarkGray}{rgb}{0.66,0.66,0.66}
\definecolor{DarkGreen}{rgb}{0.00,0.39,0.00}
\definecolor{DarkGrey}{rgb}{0.66,0.66,0.66}
\definecolor{DarkKhaki}{rgb}{0.74,0.72,0.42}
\definecolor{DarkMagenta}{rgb}{0.55,0.00,0.55}
\definecolor{DarkOliveGreen1}{rgb}{0.79,1.00,0.44}
\definecolor{DarkOliveGreen2}{rgb}{0.74,0.93,0.41}
\definecolor{DarkOliveGreen3}{rgb}{0.64,0.80,0.35}
\definecolor{DarkOliveGreen4}{rgb}{0.43,0.55,0.24}
\definecolor{DarkOliveGreen}{rgb}{0.33,0.42,0.18}
\definecolor{DarkOrange1}{rgb}{1.00,0.50,0.00}
\definecolor{DarkOrange2}{rgb}{0.93,0.46,0.00}
\definecolor{DarkOrange3}{rgb}{0.80,0.40,0.00}
\definecolor{DarkOrange4}{rgb}{0.55,0.27,0.00}
\definecolor{DarkOrange}{rgb}{1.00,0.55,0.00}
\definecolor{DarkOrchid1}{rgb}{0.75,0.24,1.00}
\definecolor{DarkOrchid2}{rgb}{0.70,0.23,0.93}
\definecolor{DarkOrchid3}{rgb}{0.60,0.20,0.80}
\definecolor{DarkOrchid4}{rgb}{0.41,0.13,0.55}
\definecolor{DarkOrchid}{rgb}{0.60,0.20,0.80}
\definecolor{DarkRed}{rgb}{0.55,0.00,0.00}
\definecolor{DarkSalmon}{rgb}{0.91,0.59,0.48}
\definecolor{DarkSeaGreen1}{rgb}{0.76,1.00,0.76}
\definecolor{DarkSeaGreen2}{rgb}{0.71,0.93,0.71}
\definecolor{DarkSeaGreen3}{rgb}{0.61,0.80,0.61}
\definecolor{DarkSeaGreen4}{rgb}{0.41,0.55,0.41}
\definecolor{DarkSeaGreen}{rgb}{0.56,0.74,0.56}
\definecolor{DarkSlateBlue}{rgb}{0.28,0.24,0.55}
\definecolor{DarkSlateGray1}{rgb}{0.59,1.00,1.00}
\definecolor{DarkSlateGray2}{rgb}{0.55,0.93,0.93}
\definecolor{DarkSlateGray3}{rgb}{0.47,0.80,0.80}
\definecolor{DarkSlateGray4}{rgb}{0.32,0.55,0.55}
\definecolor{DarkSlateGray}{rgb}{0.18,0.31,0.31}
\definecolor{DarkSlateGrey}{rgb}{0.18,0.31,0.31}
\definecolor{DarkTurquoise}{rgb}{0.00,0.81,0.82}
\definecolor{DarkViolet}{rgb}{0.58,0.00,0.83}
\definecolor{DeepPink1}{rgb}{1.00,0.08,0.58}
\definecolor{DeepPink2}{rgb}{0.93,0.07,0.54}
\definecolor{DeepPink3}{rgb}{0.80,0.06,0.46}
\definecolor{DeepPink4}{rgb}{0.55,0.04,0.31}
\definecolor{DeepPink}{rgb}{1.00,0.08,0.58}
\definecolor{DeepSkyBlue1}{rgb}{0.00,0.75,1.00}
\definecolor{DeepSkyBlue2}{rgb}{0.00,0.70,0.93}
\definecolor{DeepSkyBlue3}{rgb}{0.00,0.60,0.80}
\definecolor{DeepSkyBlue4}{rgb}{0.00,0.41,0.55}
\definecolor{DeepSkyBlue}{rgb}{0.00,0.75,1.00}
\definecolor{DimGray}{rgb}{0.41,0.41,0.41}
\definecolor{DimGrey}{rgb}{0.41,0.41,0.41}
\definecolor{DodgerBlue1}{rgb}{0.12,0.56,1.00}
\definecolor{DodgerBlue2}{rgb}{0.11,0.53,0.93}
\definecolor{DodgerBlue3}{rgb}{0.09,0.45,0.80}
\definecolor{DodgerBlue4}{rgb}{0.06,0.31,0.55}
\definecolor{DodgerBlue}{rgb}{0.12,0.56,1.00}
\definecolor{FloralWhite}{rgb}{1.00,0.98,0.94}
\definecolor{ForestGreen}{rgb}{0.13,0.55,0.13}
\definecolor{GhostWhite}{rgb}{0.97,0.97,1.00}
\definecolor{GreenYellow}{rgb}{0.68,1.00,0.18}
\definecolor{HotPink1}{rgb}{1.00,0.43,0.71}
\definecolor{HotPink2}{rgb}{0.93,0.42,0.65}
\definecolor{HotPink3}{rgb}{0.80,0.38,0.56}
\definecolor{HotPink4}{rgb}{0.55,0.23,0.38}
\definecolor{HotPink}{rgb}{1.00,0.41,0.71}
\definecolor{IndianRed1}{rgb}{1.00,0.42,0.42}
\definecolor{IndianRed2}{rgb}{0.93,0.39,0.39}
\definecolor{IndianRed3}{rgb}{0.80,0.33,0.33}
\definecolor{IndianRed4}{rgb}{0.55,0.23,0.23}
\definecolor{IndianRed}{rgb}{0.80,0.36,0.36}
\definecolor{LavenderBlush1}{rgb}{1.00,0.94,0.96}
\definecolor{LavenderBlush2}{rgb}{0.93,0.88,0.90}
\definecolor{LavenderBlush3}{rgb}{0.80,0.76,0.77}
\definecolor{LavenderBlush4}{rgb}{0.55,0.51,0.53}
\definecolor{LavenderBlush}{rgb}{1.00,0.94,0.96}
\definecolor{LawnGreen}{rgb}{0.49,0.99,0.00}
\definecolor{LemonChiffon1}{rgb}{1.00,0.98,0.80}
\definecolor{LemonChiffon2}{rgb}{0.93,0.91,0.75}
\definecolor{LemonChiffon3}{rgb}{0.80,0.79,0.65}
\definecolor{LemonChiffon4}{rgb}{0.55,0.54,0.44}
\definecolor{LemonChiffon}{rgb}{1.00,0.98,0.80}
\definecolor{LightBlue1}{rgb}{0.75,0.94,1.00}
\definecolor{LightBlue2}{rgb}{0.70,0.87,0.93}
\definecolor{LightBlue3}{rgb}{0.60,0.75,0.80}
\definecolor{LightBlue4}{rgb}{0.41,0.51,0.55}
\definecolor{LightBlue}{rgb}{0.68,0.85,0.90}
\definecolor{LightCoral}{rgb}{0.94,0.50,0.50}
\definecolor{LightCyan1}{rgb}{0.88,1.00,1.00}
\definecolor{LightCyan2}{rgb}{0.82,0.93,0.93}
\definecolor{LightCyan3}{rgb}{0.71,0.80,0.80}
\definecolor{LightCyan4}{rgb}{0.48,0.55,0.55}
\definecolor{LightCyan}{rgb}{0.88,1.00,1.00}
\definecolor{LightGoldenrod1}{rgb}{1.00,0.93,0.55}
\definecolor{LightGoldenrod2}{rgb}{0.93,0.86,0.51}
\definecolor{LightGoldenrod3}{rgb}{0.80,0.75,0.44}
\definecolor{LightGoldenrod4}{rgb}{0.55,0.51,0.30}
\definecolor{LightGoldenrodYellow}{rgb}{0.98,0.98,0.82}
\definecolor{LightGoldenrod}{rgb}{0.93,0.87,0.51}
\definecolor{LightGray}{rgb}{0.83,0.83,0.83}
\definecolor{LightGreen}{rgb}{0.56,0.93,0.56}
\definecolor{LightGrey}{rgb}{0.83,0.83,0.83}
\definecolor{LightPink1}{rgb}{1.00,0.68,0.73}
\definecolor{LightPink2}{rgb}{0.93,0.64,0.68}
\definecolor{LightPink3}{rgb}{0.80,0.55,0.58}
\definecolor{LightPink4}{rgb}{0.55,0.37,0.40}
\definecolor{LightPink}{rgb}{1.00,0.71,0.76}
\definecolor{LightSalmon1}{rgb}{1.00,0.63,0.48}
\definecolor{LightSalmon2}{rgb}{0.93,0.58,0.45}
\definecolor{LightSalmon3}{rgb}{0.80,0.51,0.38}
\definecolor{LightSalmon4}{rgb}{0.55,0.34,0.26}
\definecolor{LightSalmon}{rgb}{1.00,0.63,0.48}
\definecolor{LightSeaGreen}{rgb}{0.13,0.70,0.67}
\definecolor{LightSkyBlue1}{rgb}{0.69,0.89,1.00}
\definecolor{LightSkyBlue2}{rgb}{0.64,0.83,0.93}
\definecolor{LightSkyBlue3}{rgb}{0.55,0.71,0.80}
\definecolor{LightSkyBlue4}{rgb}{0.38,0.48,0.55}
\definecolor{LightSkyBlue}{rgb}{0.53,0.81,0.98}
\definecolor{LightSlateBlue}{rgb}{0.52,0.44,1.00}
\definecolor{LightSlateGray}{rgb}{0.47,0.53,0.60}
\definecolor{LightSlateGrey}{rgb}{0.47,0.53,0.60}
\definecolor{LightSteelBlue1}{rgb}{0.79,0.88,1.00}
\definecolor{LightSteelBlue2}{rgb}{0.74,0.82,0.93}
\definecolor{LightSteelBlue3}{rgb}{0.64,0.71,0.80}
\definecolor{LightSteelBlue4}{rgb}{0.43,0.48,0.55}
\definecolor{LightSteelBlue}{rgb}{0.69,0.77,0.87}
\definecolor{LightYellow1}{rgb}{1.00,1.00,0.88}
\definecolor{LightYellow2}{rgb}{0.93,0.93,0.82}
\definecolor{LightYellow3}{rgb}{0.80,0.80,0.71}
\definecolor{LightYellow4}{rgb}{0.55,0.55,0.48}
\definecolor{LightYellow}{rgb}{1.00,1.00,0.88}
\definecolor{LimeGreen}{rgb}{0.20,0.80,0.20}
\definecolor{MediumAquamarine}{rgb}{0.40,0.80,0.67}
\definecolor{MediumBlue}{rgb}{0.00,0.00,0.80}
\definecolor{MediumOrchid1}{rgb}{0.88,0.40,1.00}
\definecolor{MediumOrchid2}{rgb}{0.82,0.37,0.93}
\definecolor{MediumOrchid3}{rgb}{0.71,0.32,0.80}
\definecolor{MediumOrchid4}{rgb}{0.48,0.22,0.55}
\definecolor{MediumOrchid}{rgb}{0.73,0.33,0.83}
\definecolor{MediumPurple1}{rgb}{0.67,0.51,1.00}
\definecolor{MediumPurple2}{rgb}{0.62,0.47,0.93}
\definecolor{MediumPurple3}{rgb}{0.54,0.41,0.80}
\definecolor{MediumPurple4}{rgb}{0.36,0.28,0.55}
\definecolor{MediumPurple}{rgb}{0.58,0.44,0.86}
\definecolor{MediumSeaGreen}{rgb}{0.24,0.70,0.44}
\definecolor{MediumSlateBlue}{rgb}{0.48,0.41,0.93}
\definecolor{MediumSpringGreen}{rgb}{0.00,0.98,0.60}
\definecolor{MediumTurquoise}{rgb}{0.28,0.82,0.80}
\definecolor{MediumVioletRed}{rgb}{0.78,0.08,0.52}
\definecolor{MidnightBlue}{rgb}{0.10,0.10,0.44}
\definecolor{MintCream}{rgb}{0.96,1.00,0.98}
\definecolor{MistyRose1}{rgb}{1.00,0.89,0.88}
\definecolor{MistyRose2}{rgb}{0.93,0.84,0.82}
\definecolor{MistyRose3}{rgb}{0.80,0.72,0.71}
\definecolor{MistyRose4}{rgb}{0.55,0.49,0.48}
\definecolor{MistyRose}{rgb}{1.00,0.89,0.88}
\definecolor{NavajoWhite1}{rgb}{1.00,0.87,0.68}
\definecolor{NavajoWhite2}{rgb}{0.93,0.81,0.63}
\definecolor{NavajoWhite3}{rgb}{0.80,0.70,0.55}
\definecolor{NavajoWhite4}{rgb}{0.55,0.47,0.37}
\definecolor{NavajoWhite}{rgb}{1.00,0.87,0.68}
\definecolor{NavyBlue}{rgb}{0.00,0.00,0.50}
\definecolor{OldLace}{rgb}{0.99,0.96,0.90}
\definecolor{OliveDrab1}{rgb}{0.75,1.00,0.24}
\definecolor{OliveDrab2}{rgb}{0.70,0.93,0.23}
\definecolor{OliveDrab3}{rgb}{0.60,0.80,0.20}
\definecolor{OliveDrab4}{rgb}{0.41,0.55,0.13}
\definecolor{OliveDrab}{rgb}{0.42,0.56,0.14}
\definecolor{OrangeRed1}{rgb}{1.00,0.27,0.00}
\definecolor{OrangeRed2}{rgb}{0.93,0.25,0.00}
\definecolor{OrangeRed3}{rgb}{0.80,0.22,0.00}
\definecolor{OrangeRed4}{rgb}{0.55,0.15,0.00}
\definecolor{OrangeRed}{rgb}{1.00,0.27,0.00}
\definecolor{PaleGoldenrod}{rgb}{0.93,0.91,0.67}
\definecolor{PaleGreen1}{rgb}{0.60,1.00,0.60}
\definecolor{PaleGreen2}{rgb}{0.56,0.93,0.56}
\definecolor{PaleGreen3}{rgb}{0.49,0.80,0.49}
\definecolor{PaleGreen4}{rgb}{0.33,0.55,0.33}
\definecolor{PaleGreen}{rgb}{0.60,0.98,0.60}
\definecolor{PaleTurquoise1}{rgb}{0.73,1.00,1.00}
\definecolor{PaleTurquoise2}{rgb}{0.68,0.93,0.93}
\definecolor{PaleTurquoise3}{rgb}{0.59,0.80,0.80}
\definecolor{PaleTurquoise4}{rgb}{0.40,0.55,0.55}
\definecolor{PaleTurquoise}{rgb}{0.69,0.93,0.93}
\definecolor{PaleVioletRed1}{rgb}{1.00,0.51,0.67}
\definecolor{PaleVioletRed2}{rgb}{0.93,0.47,0.62}
\definecolor{PaleVioletRed3}{rgb}{0.80,0.41,0.54}
\definecolor{PaleVioletRed4}{rgb}{0.55,0.28,0.36}
\definecolor{PaleVioletRed}{rgb}{0.86,0.44,0.58}
\definecolor{PapayaWhip}{rgb}{1.00,0.94,0.84}
\definecolor{PeachPuff1}{rgb}{1.00,0.85,0.73}
\definecolor{PeachPuff2}{rgb}{0.93,0.80,0.68}
\definecolor{PeachPuff3}{rgb}{0.80,0.69,0.58}
\definecolor{PeachPuff4}{rgb}{0.55,0.47,0.40}
\definecolor{PeachPuff}{rgb}{1.00,0.85,0.73}
\definecolor{PowderBlue}{rgb}{0.69,0.88,0.90}
\definecolor{RosyBrown1}{rgb}{1.00,0.76,0.76}
\definecolor{RosyBrown2}{rgb}{0.93,0.71,0.71}
\definecolor{RosyBrown3}{rgb}{0.80,0.61,0.61}
\definecolor{RosyBrown4}{rgb}{0.55,0.41,0.41}
\definecolor{RosyBrown}{rgb}{0.74,0.56,0.56}
\definecolor{RoyalBlue1}{rgb}{0.28,0.46,1.00}
\definecolor{RoyalBlue2}{rgb}{0.26,0.43,0.93}
\definecolor{RoyalBlue3}{rgb}{0.23,0.37,0.80}
\definecolor{RoyalBlue4}{rgb}{0.15,0.25,0.55}
\definecolor{RoyalBlue}{rgb}{0.25,0.41,0.88}
\definecolor{SaddleBrown}{rgb}{0.55,0.27,0.07}
\definecolor{SandyBrown}{rgb}{0.96,0.64,0.38}
\definecolor{SeaGreen1}{rgb}{0.33,1.00,0.62}
\definecolor{SeaGreen2}{rgb}{0.31,0.93,0.58}
\definecolor{SeaGreen3}{rgb}{0.26,0.80,0.50}
\definecolor{SeaGreen4}{rgb}{0.18,0.55,0.34}
\definecolor{SeaGreen}{rgb}{0.18,0.55,0.34}
\definecolor{SkyBlue1}{rgb}{0.53,0.81,1.00}
\definecolor{SkyBlue2}{rgb}{0.49,0.75,0.93}
\definecolor{SkyBlue3}{rgb}{0.42,0.65,0.80}
\definecolor{SkyBlue4}{rgb}{0.29,0.44,0.55}
\definecolor{SkyBlue}{rgb}{0.53,0.81,0.92}
\definecolor{SlateBlue1}{rgb}{0.51,0.44,1.00}
\definecolor{SlateBlue2}{rgb}{0.48,0.40,0.93}
\definecolor{SlateBlue3}{rgb}{0.41,0.35,0.80}
\definecolor{SlateBlue4}{rgb}{0.28,0.24,0.55}
\definecolor{SlateBlue}{rgb}{0.42,0.35,0.80}
\definecolor{SlateGray1}{rgb}{0.78,0.89,1.00}
\definecolor{SlateGray2}{rgb}{0.73,0.83,0.93}
\definecolor{SlateGray3}{rgb}{0.62,0.71,0.80}
\definecolor{SlateGray4}{rgb}{0.42,0.48,0.55}
\definecolor{SlateGray}{rgb}{0.44,0.50,0.56}
\definecolor{SlateGrey}{rgb}{0.44,0.50,0.56}
\definecolor{SpringGreen1}{rgb}{0.00,1.00,0.50}
\definecolor{SpringGreen2}{rgb}{0.00,0.93,0.46}
\definecolor{SpringGreen3}{rgb}{0.00,0.80,0.40}
\definecolor{SpringGreen4}{rgb}{0.00,0.55,0.27}
\definecolor{SpringGreen}{rgb}{0.00,1.00,0.50}
\definecolor{SteelBlue1}{rgb}{0.39,0.72,1.00}
\definecolor{SteelBlue2}{rgb}{0.36,0.67,0.93}
\definecolor{SteelBlue3}{rgb}{0.31,0.58,0.80}
\definecolor{SteelBlue4}{rgb}{0.21,0.39,0.55}
\definecolor{SteelBlue}{rgb}{0.27,0.51,0.71}
\definecolor{VioletRed1}{rgb}{1.00,0.24,0.59}
\definecolor{VioletRed2}{rgb}{0.93,0.23,0.55}
\definecolor{VioletRed3}{rgb}{0.80,0.20,0.47}
\definecolor{VioletRed4}{rgb}{0.55,0.13,0.32}
\definecolor{VioletRed}{rgb}{0.82,0.13,0.56}
\definecolor{WhiteSmoke}{rgb}{0.96,0.96,0.96}
\definecolor{YellowGreen}{rgb}{0.60,0.80,0.20}
\definecolor{aliceblue}{rgb}{0.94,0.97,1.00}
\definecolor{antiquewhite}{rgb}{0.98,0.92,0.84}
\definecolor{aquamarine1}{rgb}{0.50,1.00,0.83}
\definecolor{aquamarine2}{rgb}{0.46,0.93,0.78}
\definecolor{aquamarine3}{rgb}{0.40,0.80,0.67}
\definecolor{aquamarine4}{rgb}{0.27,0.55,0.45}
\definecolor{aquamarine}{rgb}{0.50,1.00,0.83}
\definecolor{azure1}{rgb}{0.94,1.00,1.00}
\definecolor{azure2}{rgb}{0.88,0.93,0.93}
\definecolor{azure3}{rgb}{0.76,0.80,0.80}
\definecolor{azure4}{rgb}{0.51,0.55,0.55}
\definecolor{azure}{rgb}{0.94,1.00,1.00}
\definecolor{beige}{rgb}{0.96,0.96,0.86}
\definecolor{bisque1}{rgb}{1.00,0.89,0.77}
\definecolor{bisque2}{rgb}{0.93,0.84,0.72}
\definecolor{bisque3}{rgb}{0.80,0.72,0.62}
\definecolor{bisque4}{rgb}{0.55,0.49,0.42}
\definecolor{bisque}{rgb}{1.00,0.89,0.77}
\definecolor{black}{rgb}{0.00,0.00,0.00}
\definecolor{blanchedalmond}{rgb}{1.00,0.92,0.80}
\definecolor{blue1}{rgb}{0.00,0.00,1.00}
\definecolor{blue2}{rgb}{0.00,0.00,0.93}
\definecolor{blue3}{rgb}{0.00,0.00,0.80}
\definecolor{blue4}{rgb}{0.00,0.00,0.55}
\definecolor{blueviolet}{rgb}{0.54,0.17,0.89}
\definecolor{blue}{rgb}{0.00,0.00,1.00}
\definecolor{brown1}{rgb}{1.00,0.25,0.25}
\definecolor{brown2}{rgb}{0.93,0.23,0.23}
\definecolor{brown3}{rgb}{0.80,0.20,0.20}
\definecolor{brown4}{rgb}{0.55,0.14,0.14}
\definecolor{brown}{rgb}{0.65,0.16,0.16}
\definecolor{burlywood1}{rgb}{1.00,0.83,0.61}
\definecolor{burlywood2}{rgb}{0.93,0.77,0.57}
\definecolor{burlywood3}{rgb}{0.80,0.67,0.49}
\definecolor{burlywood4}{rgb}{0.55,0.45,0.33}
\definecolor{burlywood}{rgb}{0.87,0.72,0.53}
\definecolor{cadetblue}{rgb}{0.37,0.62,0.63}
\definecolor{chartreuse1}{rgb}{0.50,1.00,0.00}
\definecolor{chartreuse2}{rgb}{0.46,0.93,0.00}
\definecolor{chartreuse3}{rgb}{0.40,0.80,0.00}
\definecolor{chartreuse4}{rgb}{0.27,0.55,0.00}
\definecolor{chartreuse}{rgb}{0.50,1.00,0.00}
\definecolor{chocolate1}{rgb}{1.00,0.50,0.14}
\definecolor{chocolate2}{rgb}{0.93,0.46,0.13}
\definecolor{chocolate3}{rgb}{0.80,0.40,0.11}
\definecolor{chocolate4}{rgb}{0.55,0.27,0.07}
\definecolor{chocolate}{rgb}{0.82,0.41,0.12}
\definecolor{coral1}{rgb}{1.00,0.45,0.34}
\definecolor{coral2}{rgb}{0.93,0.42,0.31}
\definecolor{coral3}{rgb}{0.80,0.36,0.27}
\definecolor{coral4}{rgb}{0.55,0.24,0.18}
\definecolor{coral}{rgb}{1.00,0.50,0.31}
\definecolor{cornflowerblue}{rgb}{0.39,0.58,0.93}
\definecolor{cornsilk1}{rgb}{1.00,0.97,0.86}
\definecolor{cornsilk2}{rgb}{0.93,0.91,0.80}
\definecolor{cornsilk3}{rgb}{0.80,0.78,0.69}
\definecolor{cornsilk4}{rgb}{0.55,0.53,0.47}
\definecolor{cornsilk}{rgb}{1.00,0.97,0.86}
\definecolor{cyan1}{rgb}{0.00,1.00,1.00}
\definecolor{cyan2}{rgb}{0.00,0.93,0.93}
\definecolor{cyan3}{rgb}{0.00,0.80,0.80}
\definecolor{cyan4}{rgb}{0.00,0.55,0.55}
\definecolor{cyan}{rgb}{0.00,1.00,1.00}
\definecolor{darkblue}{rgb}{0.00,0.00,0.55}
\definecolor{darkcyan}{rgb}{0.00,0.55,0.55}
\definecolor{darkgoldenrod}{rgb}{0.72,0.53,0.04}
\definecolor{darkgray}{rgb}{0.66,0.66,0.66}
\definecolor{darkgreen}{rgb}{0.00,0.39,0.00}
\definecolor{darkgrey}{rgb}{0.66,0.66,0.66}
\definecolor{darkkhaki}{rgb}{0.74,0.72,0.42}
\definecolor{darkmagenta}{rgb}{0.55,0.00,0.55}
\definecolor{darkolive}{rgb}{0.33,0.42,0.18}
\definecolor{darkorange}{rgb}{1.00,0.55,0.00}
\definecolor{darkorchid}{rgb}{0.60,0.20,0.80}
\definecolor{darkred}{rgb}{0.55,0.00,0.00}
\definecolor{darksalmon}{rgb}{0.91,0.59,0.48}
\definecolor{darksea}{rgb}{0.56,0.74,0.56}
\definecolor{darkslate}{rgb}{0.18,0.31,0.31}
\definecolor{darkslate}{rgb}{0.18,0.31,0.31}
\definecolor{darkslate}{rgb}{0.28,0.24,0.55}
\definecolor{darkturquoise}{rgb}{0.00,0.81,0.82}
\definecolor{darkviolet}{rgb}{0.58,0.00,0.83}
\definecolor{deeppink}{rgb}{1.00,0.08,0.58}
\definecolor{deepsky}{rgb}{0.00,0.75,1.00}
\definecolor{dimgray}{rgb}{0.41,0.41,0.41}
\definecolor{dimgrey}{rgb}{0.41,0.41,0.41}
\definecolor{dodgerblue}{rgb}{0.12,0.56,1.00}
\definecolor{firebrick1}{rgb}{1.00,0.19,0.19}
\definecolor{firebrick2}{rgb}{0.93,0.17,0.17}
\definecolor{firebrick3}{rgb}{0.80,0.15,0.15}
\definecolor{firebrick4}{rgb}{0.55,0.10,0.10}
\definecolor{firebrick}{rgb}{0.70,0.13,0.13}
\definecolor{floralwhite}{rgb}{1.00,0.98,0.94}
\definecolor{forestgreen}{rgb}{0.13,0.55,0.13}
\definecolor{gainsboro}{rgb}{0.86,0.86,0.86}
\definecolor{ghostwhite}{rgb}{0.97,0.97,1.00}
\definecolor{gold1}{rgb}{1.00,0.84,0.00}
\definecolor{gold2}{rgb}{0.93,0.79,0.00}
\definecolor{gold3}{rgb}{0.80,0.68,0.00}
\definecolor{gold4}{rgb}{0.55,0.46,0.00}
\definecolor{goldenrod1}{rgb}{1.00,0.76,0.15}
\definecolor{goldenrod2}{rgb}{0.93,0.71,0.13}
\definecolor{goldenrod3}{rgb}{0.80,0.61,0.11}
\definecolor{goldenrod4}{rgb}{0.55,0.41,0.08}
\definecolor{goldenrod}{rgb}{0.85,0.65,0.13}
\definecolor{gold}{rgb}{1.00,0.84,0.00}
\definecolor{gray0}{rgb}{0.00,0.00,0.00}
\definecolor{gray100}{rgb}{1.00,1.00,1.00}
\definecolor{gray10}{rgb}{0.10,0.10,0.10}
\definecolor{gray11}{rgb}{0.11,0.11,0.11}
\definecolor{gray12}{rgb}{0.12,0.12,0.12}
\definecolor{gray13}{rgb}{0.13,0.13,0.13}
\definecolor{gray14}{rgb}{0.14,0.14,0.14}
\definecolor{gray15}{rgb}{0.15,0.15,0.15}
\definecolor{gray16}{rgb}{0.16,0.16,0.16}
\definecolor{gray17}{rgb}{0.17,0.17,0.17}
\definecolor{gray18}{rgb}{0.18,0.18,0.18}
\definecolor{gray19}{rgb}{0.19,0.19,0.19}
\definecolor{gray1}{rgb}{0.01,0.01,0.01}
\definecolor{gray20}{rgb}{0.20,0.20,0.20}
\definecolor{gray21}{rgb}{0.21,0.21,0.21}
\definecolor{gray22}{rgb}{0.22,0.22,0.22}
\definecolor{gray23}{rgb}{0.23,0.23,0.23}
\definecolor{gray24}{rgb}{0.24,0.24,0.24}
\definecolor{gray25}{rgb}{0.25,0.25,0.25}
\definecolor{gray26}{rgb}{0.26,0.26,0.26}
\definecolor{gray27}{rgb}{0.27,0.27,0.27}
\definecolor{gray28}{rgb}{0.28,0.28,0.28}
\definecolor{gray29}{rgb}{0.29,0.29,0.29}
\definecolor{gray2}{rgb}{0.02,0.02,0.02}
\definecolor{gray30}{rgb}{0.30,0.30,0.30}
\definecolor{gray31}{rgb}{0.31,0.31,0.31}
\definecolor{gray32}{rgb}{0.32,0.32,0.32}
\definecolor{gray33}{rgb}{0.33,0.33,0.33}
\definecolor{gray34}{rgb}{0.34,0.34,0.34}
\definecolor{gray35}{rgb}{0.35,0.35,0.35}
\definecolor{gray36}{rgb}{0.36,0.36,0.36}
\definecolor{gray37}{rgb}{0.37,0.37,0.37}
\definecolor{gray38}{rgb}{0.38,0.38,0.38}
\definecolor{gray39}{rgb}{0.39,0.39,0.39}
\definecolor{gray3}{rgb}{0.03,0.03,0.03}
\definecolor{gray40}{rgb}{0.40,0.40,0.40}
\definecolor{gray41}{rgb}{0.41,0.41,0.41}
\definecolor{gray42}{rgb}{0.42,0.42,0.42}
\definecolor{gray43}{rgb}{0.43,0.43,0.43}
\definecolor{gray44}{rgb}{0.44,0.44,0.44}
\definecolor{gray45}{rgb}{0.45,0.45,0.45}
\definecolor{gray46}{rgb}{0.46,0.46,0.46}
\definecolor{gray47}{rgb}{0.47,0.47,0.47}
\definecolor{gray48}{rgb}{0.48,0.48,0.48}
\definecolor{gray49}{rgb}{0.49,0.49,0.49}
\definecolor{gray4}{rgb}{0.04,0.04,0.04}
\definecolor{gray50}{rgb}{0.50,0.50,0.50}
\definecolor{gray51}{rgb}{0.51,0.51,0.51}
\definecolor{gray52}{rgb}{0.52,0.52,0.52}
\definecolor{gray53}{rgb}{0.53,0.53,0.53}
\definecolor{gray54}{rgb}{0.54,0.54,0.54}
\definecolor{gray55}{rgb}{0.55,0.55,0.55}
\definecolor{gray56}{rgb}{0.56,0.56,0.56}
\definecolor{gray57}{rgb}{0.57,0.57,0.57}
\definecolor{gray58}{rgb}{0.58,0.58,0.58}
\definecolor{gray59}{rgb}{0.59,0.59,0.59}
\definecolor{gray5}{rgb}{0.05,0.05,0.05}
\definecolor{gray60}{rgb}{0.60,0.60,0.60}
\definecolor{gray61}{rgb}{0.61,0.61,0.61}
\definecolor{gray62}{rgb}{0.62,0.62,0.62}
\definecolor{gray63}{rgb}{0.63,0.63,0.63}
\definecolor{gray64}{rgb}{0.64,0.64,0.64}
\definecolor{gray65}{rgb}{0.65,0.65,0.65}
\definecolor{gray66}{rgb}{0.66,0.66,0.66}
\definecolor{gray67}{rgb}{0.67,0.67,0.67}
\definecolor{gray68}{rgb}{0.68,0.68,0.68}
\definecolor{gray69}{rgb}{0.69,0.69,0.69}
\definecolor{gray6}{rgb}{0.06,0.06,0.06}
\definecolor{gray70}{rgb}{0.70,0.70,0.70}
\definecolor{gray71}{rgb}{0.71,0.71,0.71}
\definecolor{gray72}{rgb}{0.72,0.72,0.72}
\definecolor{gray73}{rgb}{0.73,0.73,0.73}
\definecolor{gray74}{rgb}{0.74,0.74,0.74}
\definecolor{gray75}{rgb}{0.75,0.75,0.75}
\definecolor{gray76}{rgb}{0.76,0.76,0.76}
\definecolor{gray77}{rgb}{0.77,0.77,0.77}
\definecolor{gray78}{rgb}{0.78,0.78,0.78}
\definecolor{gray79}{rgb}{0.79,0.79,0.79}
\definecolor{gray7}{rgb}{0.07,0.07,0.07}
\definecolor{gray80}{rgb}{0.80,0.80,0.80}
\definecolor{gray81}{rgb}{0.81,0.81,0.81}
\definecolor{gray82}{rgb}{0.82,0.82,0.82}
\definecolor{gray83}{rgb}{0.83,0.83,0.83}
\definecolor{gray84}{rgb}{0.84,0.84,0.84}
\definecolor{gray85}{rgb}{0.85,0.85,0.85}
\definecolor{gray86}{rgb}{0.86,0.86,0.86}
\definecolor{gray87}{rgb}{0.87,0.87,0.87}
\definecolor{gray88}{rgb}{0.88,0.88,0.88}
\definecolor{gray89}{rgb}{0.89,0.89,0.89}
\definecolor{gray8}{rgb}{0.08,0.08,0.08}
\definecolor{gray90}{rgb}{0.90,0.90,0.90}
\definecolor{gray91}{rgb}{0.91,0.91,0.91}
\definecolor{gray92}{rgb}{0.92,0.92,0.92}
\definecolor{gray93}{rgb}{0.93,0.93,0.93}
\definecolor{gray94}{rgb}{0.94,0.94,0.94}
\definecolor{gray95}{rgb}{0.95,0.95,0.95}
\definecolor{gray96}{rgb}{0.96,0.96,0.96}
\definecolor{gray97}{rgb}{0.97,0.97,0.97}
\definecolor{gray98}{rgb}{0.98,0.98,0.98}
\definecolor{gray99}{rgb}{0.99,0.99,0.99}
\definecolor{gray9}{rgb}{0.09,0.09,0.09}
\definecolor{gray}{rgb}{0.75,0.75,0.75}
\definecolor{green1}{rgb}{0.00,1.00,0.00}
\definecolor{green2}{rgb}{0.00,0.93,0.00}
\definecolor{green3}{rgb}{0.00,0.80,0.00}
\definecolor{green4}{rgb}{0.00,0.55,0.00}
\definecolor{greenyellow}{rgb}{0.68,1.00,0.18}
\definecolor{green}{rgb}{0.00,1.00,0.00}
\definecolor{grey0}{rgb}{0.00,0.00,0.00}
\definecolor{grey100}{rgb}{1.00,1.00,1.00}
\definecolor{grey10}{rgb}{0.10,0.10,0.10}
\definecolor{grey11}{rgb}{0.11,0.11,0.11}
\definecolor{grey12}{rgb}{0.12,0.12,0.12}
\definecolor{grey13}{rgb}{0.13,0.13,0.13}
\definecolor{grey14}{rgb}{0.14,0.14,0.14}
\definecolor{grey15}{rgb}{0.15,0.15,0.15}
\definecolor{grey16}{rgb}{0.16,0.16,0.16}
\definecolor{grey17}{rgb}{0.17,0.17,0.17}
\definecolor{grey18}{rgb}{0.18,0.18,0.18}
\definecolor{grey19}{rgb}{0.19,0.19,0.19}
\definecolor{grey1}{rgb}{0.01,0.01,0.01}
\definecolor{grey20}{rgb}{0.20,0.20,0.20}
\definecolor{grey21}{rgb}{0.21,0.21,0.21}
\definecolor{grey22}{rgb}{0.22,0.22,0.22}
\definecolor{grey23}{rgb}{0.23,0.23,0.23}
\definecolor{grey24}{rgb}{0.24,0.24,0.24}
\definecolor{grey25}{rgb}{0.25,0.25,0.25}
\definecolor{grey26}{rgb}{0.26,0.26,0.26}
\definecolor{grey27}{rgb}{0.27,0.27,0.27}
\definecolor{grey28}{rgb}{0.28,0.28,0.28}
\definecolor{grey29}{rgb}{0.29,0.29,0.29}
\definecolor{grey2}{rgb}{0.02,0.02,0.02}
\definecolor{grey30}{rgb}{0.30,0.30,0.30}
\definecolor{grey31}{rgb}{0.31,0.31,0.31}
\definecolor{grey32}{rgb}{0.32,0.32,0.32}
\definecolor{grey33}{rgb}{0.33,0.33,0.33}
\definecolor{grey34}{rgb}{0.34,0.34,0.34}
\definecolor{grey35}{rgb}{0.35,0.35,0.35}
\definecolor{grey36}{rgb}{0.36,0.36,0.36}
\definecolor{grey37}{rgb}{0.37,0.37,0.37}
\definecolor{grey38}{rgb}{0.38,0.38,0.38}
\definecolor{grey39}{rgb}{0.39,0.39,0.39}
\definecolor{grey3}{rgb}{0.03,0.03,0.03}
\definecolor{grey40}{rgb}{0.40,0.40,0.40}
\definecolor{grey41}{rgb}{0.41,0.41,0.41}
\definecolor{grey42}{rgb}{0.42,0.42,0.42}
\definecolor{grey43}{rgb}{0.43,0.43,0.43}
\definecolor{grey44}{rgb}{0.44,0.44,0.44}
\definecolor{grey45}{rgb}{0.45,0.45,0.45}
\definecolor{grey46}{rgb}{0.46,0.46,0.46}
\definecolor{grey47}{rgb}{0.47,0.47,0.47}
\definecolor{grey48}{rgb}{0.48,0.48,0.48}
\definecolor{grey49}{rgb}{0.49,0.49,0.49}
\definecolor{grey4}{rgb}{0.04,0.04,0.04}
\definecolor{grey50}{rgb}{0.50,0.50,0.50}
\definecolor{grey51}{rgb}{0.51,0.51,0.51}
\definecolor{grey52}{rgb}{0.52,0.52,0.52}
\definecolor{grey53}{rgb}{0.53,0.53,0.53}
\definecolor{grey54}{rgb}{0.54,0.54,0.54}
\definecolor{grey55}{rgb}{0.55,0.55,0.55}
\definecolor{grey56}{rgb}{0.56,0.56,0.56}
\definecolor{grey57}{rgb}{0.57,0.57,0.57}
\definecolor{grey58}{rgb}{0.58,0.58,0.58}
\definecolor{grey59}{rgb}{0.59,0.59,0.59}
\definecolor{grey5}{rgb}{0.05,0.05,0.05}
\definecolor{grey60}{rgb}{0.60,0.60,0.60}
\definecolor{grey61}{rgb}{0.61,0.61,0.61}
\definecolor{grey62}{rgb}{0.62,0.62,0.62}
\definecolor{grey63}{rgb}{0.63,0.63,0.63}
\definecolor{grey64}{rgb}{0.64,0.64,0.64}
\definecolor{grey65}{rgb}{0.65,0.65,0.65}
\definecolor{grey66}{rgb}{0.66,0.66,0.66}
\definecolor{grey67}{rgb}{0.67,0.67,0.67}
\definecolor{grey68}{rgb}{0.68,0.68,0.68}
\definecolor{grey69}{rgb}{0.69,0.69,0.69}
\definecolor{grey6}{rgb}{0.06,0.06,0.06}
\definecolor{grey70}{rgb}{0.70,0.70,0.70}
\definecolor{grey71}{rgb}{0.71,0.71,0.71}
\definecolor{grey72}{rgb}{0.72,0.72,0.72}
\definecolor{grey73}{rgb}{0.73,0.73,0.73}
\definecolor{grey74}{rgb}{0.74,0.74,0.74}
\definecolor{grey75}{rgb}{0.75,0.75,0.75}
\definecolor{grey76}{rgb}{0.76,0.76,0.76}
\definecolor{grey77}{rgb}{0.77,0.77,0.77}
\definecolor{grey78}{rgb}{0.78,0.78,0.78}
\definecolor{grey79}{rgb}{0.79,0.79,0.79}
\definecolor{grey7}{rgb}{0.07,0.07,0.07}
\definecolor{grey80}{rgb}{0.80,0.80,0.80}
\definecolor{grey81}{rgb}{0.81,0.81,0.81}
\definecolor{grey82}{rgb}{0.82,0.82,0.82}
\definecolor{grey83}{rgb}{0.83,0.83,0.83}
\definecolor{grey84}{rgb}{0.84,0.84,0.84}
\definecolor{grey85}{rgb}{0.85,0.85,0.85}
\definecolor{grey86}{rgb}{0.86,0.86,0.86}
\definecolor{grey87}{rgb}{0.87,0.87,0.87}
\definecolor{grey88}{rgb}{0.88,0.88,0.88}
\definecolor{grey89}{rgb}{0.89,0.89,0.89}
\definecolor{grey8}{rgb}{0.08,0.08,0.08}
\definecolor{grey90}{rgb}{0.90,0.90,0.90}
\definecolor{grey91}{rgb}{0.91,0.91,0.91}
\definecolor{grey92}{rgb}{0.92,0.92,0.92}
\definecolor{grey93}{rgb}{0.93,0.93,0.93}
\definecolor{grey94}{rgb}{0.94,0.94,0.94}
\definecolor{grey95}{rgb}{0.95,0.95,0.95}
\definecolor{grey96}{rgb}{0.96,0.96,0.96}
\definecolor{grey97}{rgb}{0.97,0.97,0.97}
\definecolor{grey98}{rgb}{0.98,0.98,0.98}
\definecolor{grey99}{rgb}{0.99,0.99,0.99}
\definecolor{grey9}{rgb}{0.09,0.09,0.09}
\definecolor{grey}{rgb}{0.75,0.75,0.75}
\definecolor{honeydew1}{rgb}{0.94,1.00,0.94}
\definecolor{honeydew2}{rgb}{0.88,0.93,0.88}
\definecolor{honeydew3}{rgb}{0.76,0.80,0.76}
\definecolor{honeydew4}{rgb}{0.51,0.55,0.51}
\definecolor{honeydew}{rgb}{0.94,1.00,0.94}
\definecolor{hotpink}{rgb}{1.00,0.41,0.71}
\definecolor{indianred}{rgb}{0.80,0.36,0.36}
\definecolor{ivory1}{rgb}{1.00,1.00,0.94}
\definecolor{ivory2}{rgb}{0.93,0.93,0.88}
\definecolor{ivory3}{rgb}{0.80,0.80,0.76}
\definecolor{ivory4}{rgb}{0.55,0.55,0.51}
\definecolor{ivory}{rgb}{1.00,1.00,0.94}
\definecolor{khaki1}{rgb}{1.00,0.96,0.56}
\definecolor{khaki2}{rgb}{0.93,0.90,0.52}
\definecolor{khaki3}{rgb}{0.80,0.78,0.45}
\definecolor{khaki4}{rgb}{0.55,0.53,0.31}
\definecolor{khaki}{rgb}{0.94,0.90,0.55}
\definecolor{lavenderblush}{rgb}{1.00,0.94,0.96}
\definecolor{lavender}{rgb}{0.90,0.90,0.98}
\definecolor{lawngreen}{rgb}{0.49,0.99,0.00}
\definecolor{lemonchiffon}{rgb}{1.00,0.98,0.80}
\definecolor{lightblue}{rgb}{0.68,0.85,0.90}
\definecolor{lightcoral}{rgb}{0.94,0.50,0.50}
\definecolor{lightcyan}{rgb}{0.88,1.00,1.00}
\definecolor{lightgoldenrod}{rgb}{0.93,0.87,0.51}
\definecolor{lightgoldenrod}{rgb}{0.98,0.98,0.82}
\definecolor{lightgray}{rgb}{0.83,0.83,0.83}
\definecolor{lightgreen}{rgb}{0.56,0.93,0.56}
\definecolor{lightgrey}{rgb}{0.83,0.83,0.83}
\definecolor{lightpink}{rgb}{1.00,0.71,0.76}
\definecolor{lightsalmon}{rgb}{1.00,0.63,0.48}
\definecolor{lightsea}{rgb}{0.13,0.70,0.67}
\definecolor{lightsky}{rgb}{0.53,0.81,0.98}
\definecolor{lightslate}{rgb}{0.47,0.53,0.60}
\definecolor{lightslate}{rgb}{0.47,0.53,0.60}
\definecolor{lightslate}{rgb}{0.52,0.44,1.00}
\definecolor{lightsteel}{rgb}{0.69,0.77,0.87}
\definecolor{lightyellow}{rgb}{1.00,1.00,0.88}
\definecolor{limegreen}{rgb}{0.20,0.80,0.20}
\definecolor{linen}{rgb}{0.98,0.94,0.90}
\definecolor{magenta1}{rgb}{1.00,0.00,1.00}
\definecolor{magenta2}{rgb}{0.93,0.00,0.93}
\definecolor{magenta3}{rgb}{0.80,0.00,0.80}
\definecolor{magenta4}{rgb}{0.55,0.00,0.55}
\definecolor{magenta}{rgb}{1.00,0.00,1.00}
\definecolor{maroon1}{rgb}{1.00,0.20,0.70}
\definecolor{maroon2}{rgb}{0.93,0.19,0.65}
\definecolor{maroon3}{rgb}{0.80,0.16,0.56}
\definecolor{maroon4}{rgb}{0.55,0.11,0.38}
\definecolor{maroon}{rgb}{0.69,0.19,0.38}
\definecolor{mediumaquamarine}{rgb}{0.40,0.80,0.67}
\definecolor{mediumblue}{rgb}{0.00,0.00,0.80}
\definecolor{mediumorchid}{rgb}{0.73,0.33,0.83}
\definecolor{mediumpurple}{rgb}{0.58,0.44,0.86}
\definecolor{mediumsea}{rgb}{0.24,0.70,0.44}
\definecolor{mediumslate}{rgb}{0.48,0.41,0.93}
\definecolor{mediumspring}{rgb}{0.00,0.98,0.60}
\definecolor{mediumturquoise}{rgb}{0.28,0.82,0.80}
\definecolor{mediumviolet}{rgb}{0.78,0.08,0.52}
\definecolor{midnightblue}{rgb}{0.10,0.10,0.44}
\definecolor{mintcream}{rgb}{0.96,1.00,0.98}
\definecolor{mistyrose}{rgb}{1.00,0.89,0.88}
\definecolor{moccasin}{rgb}{1.00,0.89,0.71}
\definecolor{navajowhite}{rgb}{1.00,0.87,0.68}
\definecolor{navyblue}{rgb}{0.00,0.00,0.50}
\definecolor{navy}{rgb}{0.00,0.00,0.50}
\definecolor{oldlace}{rgb}{0.99,0.96,0.90}
\definecolor{olivedrab}{rgb}{0.42,0.56,0.14}
\definecolor{orange1}{rgb}{1.00,0.65,0.00}
\definecolor{orange2}{rgb}{0.93,0.60,0.00}
\definecolor{orange3}{rgb}{0.80,0.52,0.00}
\definecolor{orange4}{rgb}{0.55,0.35,0.00}
\definecolor{orangered}{rgb}{1.00,0.27,0.00}
\definecolor{orange}{rgb}{1.00,0.65,0.00}
\definecolor{orchid1}{rgb}{1.00,0.51,0.98}
\definecolor{orchid2}{rgb}{0.93,0.48,0.91}
\definecolor{orchid3}{rgb}{0.80,0.41,0.79}
\definecolor{orchid4}{rgb}{0.55,0.28,0.54}
\definecolor{orchid}{rgb}{0.85,0.44,0.84}
\definecolor{palegoldenrod}{rgb}{0.93,0.91,0.67}
\definecolor{palegreen}{rgb}{0.60,0.98,0.60}
\definecolor{paleturquoise}{rgb}{0.69,0.93,0.93}
\definecolor{paleviolet}{rgb}{0.86,0.44,0.58}
\definecolor{papayawhip}{rgb}{1.00,0.94,0.84}
\definecolor{peachpuff}{rgb}{1.00,0.85,0.73}
\definecolor{peru}{rgb}{0.80,0.52,0.25}
\definecolor{pink1}{rgb}{1.00,0.71,0.77}
\definecolor{pink2}{rgb}{0.93,0.66,0.72}
\definecolor{pink3}{rgb}{0.80,0.57,0.62}
\definecolor{pink4}{rgb}{0.55,0.39,0.42}
\definecolor{pink}{rgb}{1.00,0.75,0.80}
\definecolor{plum1}{rgb}{1.00,0.73,1.00}
\definecolor{plum2}{rgb}{0.93,0.68,0.93}
\definecolor{plum3}{rgb}{0.80,0.59,0.80}
\definecolor{plum4}{rgb}{0.55,0.40,0.55}
\definecolor{plum}{rgb}{0.87,0.63,0.87}
\definecolor{powderblue}{rgb}{0.69,0.88,0.90}
\definecolor{purple1}{rgb}{0.61,0.19,1.00}
\definecolor{purple2}{rgb}{0.57,0.17,0.93}
\definecolor{purple3}{rgb}{0.49,0.15,0.80}
\definecolor{purple4}{rgb}{0.33,0.10,0.55}
\definecolor{purple}{rgb}{0.63,0.13,0.94}
\definecolor{red1}{rgb}{1.00,0.00,0.00}
\definecolor{red2}{rgb}{0.93,0.00,0.00}
\definecolor{red3}{rgb}{0.80,0.00,0.00}
\definecolor{red4}{rgb}{0.55,0.00,0.00}
\definecolor{red}{rgb}{1.00,0.00,0.00}
\definecolor{rosybrown}{rgb}{0.74,0.56,0.56}
\definecolor{royalblue}{rgb}{0.25,0.41,0.88}
\definecolor{saddlebrown}{rgb}{0.55,0.27,0.07}
\definecolor{salmon1}{rgb}{1.00,0.55,0.41}
\definecolor{salmon2}{rgb}{0.93,0.51,0.38}
\definecolor{salmon3}{rgb}{0.80,0.44,0.33}
\definecolor{salmon4}{rgb}{0.55,0.30,0.22}
\definecolor{salmon}{rgb}{0.98,0.50,0.45}
\definecolor{sandybrown}{rgb}{0.96,0.64,0.38}
\definecolor{seagreen}{rgb}{0.18,0.55,0.34}
\definecolor{seashell1}{rgb}{1.00,0.96,0.93}
\definecolor{seashell2}{rgb}{0.93,0.90,0.87}
\definecolor{seashell3}{rgb}{0.80,0.77,0.75}
\definecolor{seashell4}{rgb}{0.55,0.53,0.51}
\definecolor{seashell}{rgb}{1.00,0.96,0.93}
\definecolor{sienna1}{rgb}{1.00,0.51,0.28}
\definecolor{sienna2}{rgb}{0.93,0.47,0.26}
\definecolor{sienna3}{rgb}{0.80,0.41,0.22}
\definecolor{sienna4}{rgb}{0.55,0.28,0.15}
\definecolor{sienna}{rgb}{0.63,0.32,0.18}
\definecolor{skyblue}{rgb}{0.53,0.81,0.92}
\definecolor{slateblue}{rgb}{0.42,0.35,0.80}
\definecolor{slategray}{rgb}{0.44,0.50,0.56}
\definecolor{slategrey}{rgb}{0.44,0.50,0.56}
\definecolor{snow1}{rgb}{1.00,0.98,0.98}
\definecolor{snow2}{rgb}{0.93,0.91,0.91}
\definecolor{snow3}{rgb}{0.80,0.79,0.79}
\definecolor{snow4}{rgb}{0.55,0.54,0.54}
\definecolor{snow}{rgb}{1.00,0.98,0.98}
\definecolor{springgreen}{rgb}{0.00,1.00,0.50}
\definecolor{steelblue}{rgb}{0.27,0.51,0.71}
\definecolor{tan1}{rgb}{1.00,0.65,0.31}
\definecolor{tan2}{rgb}{0.93,0.60,0.29}
\definecolor{tan3}{rgb}{0.80,0.52,0.25}
\definecolor{tan4}{rgb}{0.55,0.35,0.17}
\definecolor{tan}{rgb}{0.82,0.71,0.55}
\definecolor{thistle1}{rgb}{1.00,0.88,1.00}
\definecolor{thistle2}{rgb}{0.93,0.82,0.93}
\definecolor{thistle3}{rgb}{0.80,0.71,0.80}
\definecolor{thistle4}{rgb}{0.55,0.48,0.55}
\definecolor{thistle}{rgb}{0.85,0.75,0.85}
\definecolor{tomato1}{rgb}{1.00,0.39,0.28}
\definecolor{tomato2}{rgb}{0.93,0.36,0.26}
\definecolor{tomato3}{rgb}{0.80,0.31,0.22}
\definecolor{tomato4}{rgb}{0.55,0.21,0.15}
\definecolor{tomato}{rgb}{1.00,0.39,0.28}
\definecolor{turquoise1}{rgb}{0.00,0.96,1.00}
\definecolor{turquoise2}{rgb}{0.00,0.90,0.93}
\definecolor{turquoise3}{rgb}{0.00,0.77,0.80}
\definecolor{turquoise4}{rgb}{0.00,0.53,0.55}
\definecolor{turquoise}{rgb}{0.25,0.88,0.82}
\definecolor{violetred}{rgb}{0.82,0.13,0.56}
\definecolor{violet}{rgb}{0.93,0.51,0.93}
\definecolor{wheat1}{rgb}{1.00,0.91,0.73}
\definecolor{wheat2}{rgb}{0.93,0.85,0.68}
\definecolor{wheat3}{rgb}{0.80,0.73,0.59}
\definecolor{wheat4}{rgb}{0.55,0.49,0.40}
\definecolor{wheat}{rgb}{0.96,0.87,0.70}
\definecolor{whitesmoke}{rgb}{0.96,0.96,0.96}
\definecolor{white}{rgb}{1.00,1.00,1.00}
\definecolor{yellow1}{rgb}{1.00,1.00,0.00}
\definecolor{yellow2}{rgb}{0.93,0.93,0.00}
\definecolor{yellow3}{rgb}{0.80,0.80,0.00}
\definecolor{yellow4}{rgb}{0.55,0.55,0.00}
\definecolor{yellowgreen}{rgb}{0.60,0.80,0.20}
\definecolor{yellow}{rgb}{1.00,1.00,0.00}

\title{Do group dynamics play a role in the evolution of member galaxies?}
\author[Hou et al.]{Annie Hou$^{1}$, Laura C. Parker$^{1}$, Michael L. Balogh$^{2}$, Sean L. McGee$^{3,4}$, David J. Wilman$^{5}$, \newauthor Jennifer L. Connelly$^{5}$, William E. Harris$^{1}$, Angus Mok$^{2}$, John S. Mulchaey$^{6}$, \newauthor Richard G. Bower$^{3}$ $\&$ Alexis Finoguenov$^{7,8}$\\
$^{1}$Department of Physics $\&$ Astronomy, McMaster University, Hamilton ON L8S 4M1, Canada\\
$^{2}$Department of Physics and Astronomy, Univeristy of Waterloo, Waterloo, Ontario, N2L 3G1, Canada\\
$^{3}$Department Of Physics, University of Durham, Durham, DH1 3LE, United Kingdom\\
$^{4}$Leiden Observatory, Leiden University, PO Box 9513, 2300 RA Leiden, the Netherlands\\
$^{5}$Max-Planck-Institut f$\ddot{u}$r Extraterrestrische Physik, Giessenbachstra\ss e,  D-85748 Garching, Germany\\
$^{6}$Observatories of the Carnegie Institution, 813 Santa Barbara Street, Pasadena, California, USA\\
$^{7}$Department of Physics, University of Helsinki, Gustaf Hallstromin katu 2a, FI-00014 Helsinki, Finlanda\\
$^{8}$CSST, University of Maryland, Baltimore County, 1000 Hilltop Circle, Baltimore, MD 21250, USA}

\begin{document}
\maketitle
\begin{abstract}
We examine galaxy groups from the present epoch to $z \sim 1$ to explore the impact of group dynamics on galaxy evolution.  We use group catalogues from the Sloan Digital Sky Survey (SDSS), the Group Environment and Evolution Collaboration (GEEC) and the high redshift GEEC2 sample to study how the observed member properties depend on galaxy stellar mass, group dynamical mass and dynamical state of the host group.  We find a strong correlation between the fraction of non-star-forming (quiescent) galaxies and galaxy stellar mass, but do not detect a significant difference in the quiescent fraction with group dynamical mass, within our sample halo mass range of $\sim 10^{13} - 10^{14.5} M_{\odot}$, or with dynamical state.  However, at $z \sim 0.4$ we do see some evidence that the quiescent fraction in low mass galaxies ($\log_{10}(M_{\rm{star}}/M_{\odot}) \lesssim 10.5$) is lower in groups with substructure.  Additionally, our results show that the fraction of groups with non-Gaussian velocity distributions increases with redshift to $z \sim 0.4$, while the amount of detected substructure remains constant to $z \sim 1$.  Based on these results, we conclude that for massive galaxies ($\log_{10}(M_{\rm{star}}/M_{\odot}) \gtrsim 10.5$), evolution is most strongly correlated to the stellar mass of a galaxy with little or no additional effect related to either the group dynamical mass or dynamical state.  For low mass galaxies, we do see some evidence of a correlation between quiescent fraction and the amount of detected substructure, highlighting the need to probe further down the stellar mass function to elucidate the role of environment in galaxy evolution.
\end{abstract}

\section{Introduction}
\label{intro}
A long-standing debate is whether the evolution of galaxies is governed primarily by internal processes (e.g.\ feedback) or those related to the external environment (e.g.\ stripping).  The morphology-density relation seen in the cores of clusters \citep{oemler74, dressler80} was one of the first observations to show that the environment may influence the properties of galaxies, where elliptical and S0 (early-type) galaxies were found preferentially in high-density regions and spiral and irregular (late-type) galaxies in low-density regions.  Since then, numerous correlations between galaxy properties and environment have been observed.  For example, differences in the distributions of colours \citep{blanton03, baldry06, hou09}, the fraction of either star-forming or quiescent galaxies \citep{kauffmann04, balogh04, wilman05b, peng10, mcgee11, patel11, sobral11, muzzin12}, and the amount of observed dust \citep{kauffmann04}.  Correlations between environment and galaxy properties appear to have been in place since at least $z \sim 1$, as the observed star formation rate (SFR)-density and specific star formation rate (SSFR = SFR/stellar mass)-density relations show variations with environment at this redshift \citep{cooper08, patel11}. 

Although there have been numerous observations of correlations between environment and galaxy properties, where red and quiescent galaxies are preferentially found in higher density regions, recent studies have suggested that internal or secular processes, traced by the mass of the galaxy, may actually be the dominant factor in galaxy evolution.  In particular, several studies have found that the properties of actively star-forming galaxies only weakly depend on the environment \citep{balogh04, wilman05b, poggianti08, peng10, tyler11}.  Similarly, \citet{muzzin12} found that although the environment does determine the fraction of galaxies that remain actively star-forming, the stellar populations of both actively star-forming and quiescent galaxies are most strongly correlated to the stellar mass of a galaxy.

The emerging picture appears to suggest that \emph{both} internal and external processes contribute to the evolution of galaxies.  Although observations have shown that stellar mass correlates well with environment, both in the local Universe \citep{hogg03, kauffmann04, blanton05, baldry06} and at $z \sim 1$ \citep{bolzonella10, sobral11}, recent studies have claimed that the effects due to the environment can still be disentangled from transformation processes traced by galaxy stellar mass \citep{peng10, sobral11, muzzin12}.  In an empirically driven picture of galaxy evolution, \citet{peng10}  claimed that the evolution of low mass galaxies ($\log_{10}(M_{\rm{star}}/M_{\odot}) \lesssim 10.5$) is dominated by environmentally driven star-formation quenching whereas high mass galaxy evolution ($\log_{10}(M_{\rm{star}}/M_{\odot}) \gtrsim 10.5$) is governed by processes which are traced by galaxy stellar mass.   

Galaxy groups are ideal for studies of the role of the environment in the evolution of galaxies.  Not only are groups the most common environment in the local Universe \citep{gh83, eke05}, but it is also believed that as many as 40 per cent of galaxies, especially low mass galaxies, that live in rich groups or clusters were pre-processed (i.e.\ had their star formation quenched) in haloes with $M_{\rm{halo}} \gtrsim 10^{13}h^{-1}M_{\odot}$ before infall \citep{mcgee09, delucia12}.

The pre-processing of galaxies in low mass groups may be driven by galaxy-galaxy interactions and mergers.  As a result of the relatively low velocity dispersion observed in groups, it has been shown that the rate of mergers is higher in the group environment with respect to both the field and richer galaxy clusters \citep{barnes85, zm98b, brough06, delucia11}.  Interactions are thought to initially trigger an intense burst of star formation \citep{sanders88, ec03, cox06, teyssier10}, which can use up the supply of cold gas and lead to the quenching of star formation, if no further gas accretes onto galaxy.  Thus, mergers and interactions can either enhance or quench star formation depending on the evolutionary stage at which the galaxy is observed.  In addition, star formation quenching in galaxies may occur as a satellite falls into a larger dark matter halo due to processes such as strangulation \citep{ltc80, balogh00, km08} and ram-pressure stripping \citep{gg72,abadi99}.  Thus, galaxy evolution appears to be related to the accretion history of the galaxy and with the number of interactions a galaxy has experienced.  By looking for correlations between group dynamics and member properties, it is possible to probe the importance of accretion history and dynamical interactions on the evolution of galaxies.

In this paper we study the dependance of galaxy evolution on galaxy stellar mass, group dynamical mass and group dynamics.  The paper is structured as follows: in Section \ref{data} we describe the data and group catalogues, as well as discuss the methods for determining stellar mass and SFR.  In Section \ref{analysis}, we look for correlations of galaxy properties with galaxy stellar mass and group dynamical mass.  In Section \ref{grpdyn}, we classify the dynamical state of our groups and compare the properties of galaxies in dynamically young and dynamically evolved systems.  We discuss our results in Section \ref{discussion} and finally present our conclusions in Section \ref{conclusion}.

Throughout this paper we assume a $\Lambda$CDM cosmology with $\Omega_{m,0} = 0.27$, $\Omega_{\Lambda,0} = 0.73$ and $H_{0} = 71$ km s$^{-1}$ Mpc$^{-1}$. 

\section{Data}
\label{data}
In order to investigate the role of group dynamics in galaxy evolution, we look at three highly complete group catalogues that span a redshift range of $0 \lesssim z \lesssim 1$.  This allows us to probe not only how the properties of the member galaxies depend on the properties of the host group, but also how these correlations evolve with redshift.  The low redshift ($0 < z < 0.12$) group sample is from the Sloan Digital Sky Survey (SDSS), the intermediate redshift ($0.15 < z < 0.55$) sample is from the Group Environment and Evolution Collaboration (GEEC) survey, and the high redshift ($0.8 < z < 1$) groups are from the GEEC2 survey (to be discussed in detail in Sections \ref{sdss}-\ref{geec2}).

\subsection{The SDSS group catalogue}
\label{sdss}
Although there are many publicly available SDSS group catalogues \citep[e.g.,][]{berlind06, yang07}, we elect to use the groups defined in \cite{mcgee11}, who applied a multi-stage approach to mimic both the observing conditions and group-finding algorithm used to identify our intermediate redshift GEEC groups (see Section \ref{geec}).  This selection allows for a better comparison of the group and galaxy properties by reducing possible effects introduced by differences in the spectroscopic completeness, limiting magnitude or in the group-finding algorithm.  
 A full description of our SDSS group catalogue is given in \citet{mcgee11}, but we give a brief summary here.  Groups were identified using galaxies observed in the SDSS Data Release 6 (DR6), which contains over 790 000 spectra in an area of $\sim$7425 deg$^{2}$ \citep{am08}.  In addition to the SDSS $ugriz$ photometry, \citet{mcgee11} made use of the overlapping GALEX Medium Imaging Survey (MIS), which covered an area of $\sim$1000 deg$^{2}$ of the SDSS \citep{martin05, morrissey07}.  The inclusion of the NUV and FUV bands is important for better estimates of the SFR.

To reproduce the observing conditions and group-finding algorithm of the second Canadian Network for Observational Cosmology (CNOC2) Galaxy Redshift Survey \citep{yee00}, on which the GEEC group catalogue is based,  \citet{mcgee11} applied the same absolute magnitude cut and then randomly removed half the remaining galaxies to match the spectroscopic completeness of the CNOC2 redshift survey (see Section 2.2).  With this sample, \citet{mcgee11} calculated the local density around each galaxy by counting the number of galaxies within a cylinder of 0.33$h_{75}^{-1}$ Mpc and a line-of-sight depth of $\pm$6.67$h_{75}^{-1}$ Mpc.  Proto-groups were then identified starting from galaxies with the highest local densities and taking all of the galaxies within the cylinder centred around the high-density galaxies as proto-group members.  Next, all of the galaxies in each of the cylinders centred on the initial members were added and the process continued until no further galaxies could be added.  Using these proto-groups, a preliminary geometric centre, redshift, velocity dispersion ($\sigma$) and virial radius ($r_{200}$: Equation \ref{r200}) were computed.  Proto-group members were then added or removed iteratively if they fell within 1.5$r_{200}$ and 3$\sigma$ of the group centre.  Once all of the proto-groups were identified, \citet{mcgee11} then added all of the SDSS galaxies back into the sample and group membership was finalized with a methodology similar to that used to identify the GEEC groups \citep{carlberg01, wilman05}.

\subsection{The GEEC group catalogue}
\label{geec}
Our intermediate redshift sample is the GEEC group catalogue, which contains $\sim$200 groups in the range of $0.1 < z < 0.55$.  The GEEC survey is based on a set of groups first identified in the CNOC2 redshift survey \citep{yee00, carlberg01}, which contained $\sim$4$\times10^{4}$ galaxies in four patches totalling $\sim$1.5 deg$^{2}$.  The original photometry was taken in the $UBVR_{c}I_{c}$ bands down to a limiting magnitude of $R_{c} = 23.0$ and spectra were obtained for more than 6000 galaxies with a completeness of 48 per cent at $R_{c} = 21.5$ \citep{yee00}.

The GEEC survey built on the CNOC2 survey by obtaining higher spectroscopic completeness to a fainter limiting magnitude with 78 per cent completeness at $R_{c} = 22$ for a subset of the groups \citep{wilman05, connelly12}.  The extensive follow-up spectroscopy was taken with LDSS2 \citep{wilman05} and IMACS \citep{connelly12} on Magellan, as well as data  from FORS2 on the Very Large Telescope \citep{connelly12}.  Additionally, we have obtained multi-wavelength data from the X-ray to the infrared (IR) observed with the following telescopes: XMM-Newton \citep{finog09}, Chandra X-ray Observatory \citep{finog09}, GALEX \citep{mcgee11}, HST-ACS \citep{wilman09}, Spitzer-MIPS \citep{tyler11}, Spitzer-IRAC \citep{wilman09b}, INGRID on the William Herschel Telescope \citep{balogh09}, and SOFI on the New Technology Telescope \citep{balogh07}.  In addition, improved optical imaging was obtained in the $ugrizBVRI$ filters from the Canada-France-Hawaii Telescope's Megacam and CFH12K imagers \citep{balogh09}. 

Group membership was defined with the  friends-of-friend (FoF) algorithm outlined in \citet{wilman05}.  Analysis of mock catalogues has shown that the contamination rate is 2.5 per cent for galaxies within $0.5 h_{75}^{-1}$ Mpc of the group centroid \citep{mcgee08}.

\subsection{The GEEC2 group catalogue}
\label{geec2}
The high redshift sample contains a subset of groups identified in the GEEC2 survey.  A detailed discussion of the GEEC2 survey is presented in \citet{balogh11} and \citet{mok13}.  The goal of the GEEC2 survey was to obtain high spectroscopic completeness for 20 galaxy groups in the redshift range of $0.8 < z < 1.0$ that were initially identified in the Cosmic Evolution Survey \citep[COSMOS -][]{scoville07} with extended X-ray emission \citep{finog07, george11}.  The follow-up spectroscopic survey is being conducted with the GMOS spectrograph on the GEMINI telescope and thus far, data have been collected for 11 of the 20 target groups with spectroscopic completeness between $\sim$0.6 and 0.75 \citep{balogh11} down to $r = 24.75$.  

\citet{balogh11} assigned group membership based on a galaxy's proximity to the measured X-ray centre.  It should be noted that although the group centroid for GEEC2 is based on X-ray emission, rather than a luminosity weighted center (used in SDSS and GEEC), \citet{connelly12} found that the difference between these two definitions is typically small ($< 18 \arcsec$) and that group membership and overall group properties did not change signficantly with either centroid.  For each group, the velocity dispersion ($\sigma$) was computed from all galaxies within 1.0 Mpc and 4000 km s$^{-1}$ of the measured group X-ray centre.  Next the \emph{rms} projected radial position from the group centroid ($R_{rms}$) was computed and all galaxies with group-centric velocities $> C_{z}\sigma$ and radial position $> C_{r}R_{rms}$ were clipped, where the typical values for $C_{z}$ and $C_{r}$ were 2.  Finally, $\sigma$ and $R_{rms}$ were re-computed and only galaxies with $z <$ 2.5$\sigma_{\rm{1Mpc}}$ and radial positions $< 2R_{rms}$ were defined as group members.

Ideally, all three group catalogues would be defined in the same way; however, an unbiased and highly complete spectroscopic survey at high redshifts is a difficult and expensive task.  Including the GEEC2 groups allows us to probe the high redshift Universe.  Additionally, GEEC2 is one of the few high redshift group catalogues with high spectroscopic completeness and more than five members per group, allowing for studies of group dynamics.  

\subsection{Spectral energy distribution fitting}
\begin{table*}
\centering
\caption{Properties of the group catalogues. \label{sedtab}}
\begin{tabular}{cccc}
\hline\hline
Catalogue & redshift range & $\#$ of groups & Photometry\\
\hline
SDSS & z $<$ 0.1 & 100 & FUV, NUV, \\
(\emph{r} $<$ 17.77) & & & \emph{u, g, r, i, z}\\
\hline
GEEC & 0.1 $<$ z $<$ 0.55 & 37 & FUV, NUV,\\
 & & & \emph{u, g, r, i, z, K}\\
(\emph{r} $<$ 23) & & & Spitzer:IRAC $\&$ MIPS\\
\hline
GEEC2 &  0.8 $<$ z $<$ 1 & 8 & FUV, NUV, U, B, V, G, \\
({r} $<$ 24.75) & & & R, I, Z, J, K \\
 & & & Spitzer:IRAC $\&$ MIPS\\
\hline
\end{tabular}
\end{table*}

\label{sed}
In Table \ref{sedtab}, we list the group catalogue, redshift range, number of groups used in this analysis and the available photometry for each sample.  We see that each of the three group catalogues has multi-wavelength data (Table \ref{sedtab}), which were used to measure stellar masses and SFRs via spectral energy distribution (SED) template-fitting.  The same fitting procedure was carried out for all catalogues from all available photometric bands.  A detailed discussion of the SED fitting procedure is given in \citet{mcgee11} for the SDSS and GEEC catalogues and in \citet{balogh11} for the GEEC2 catalogue.  The observed photometry was compared to a grid of SEDs constructed with the \citet{bc03} stellar population synthesis code for the SDSS and GEEC catalogues and the \citet{bruzual07} model for GEEC2.  A Chabrier initial mass function (IMF) was assumed for all three catalogues.  In both \citet{mcgee11} and \citet{balogh11}, the SED fitting procedure followed that outlined in \citet{salim07}, where a grid of models that uniformly sampled the allowed parameters of formation time, galaxy metallicity and a two-component dust model \citep{cf00} was created.  An exponentially declining base SFR, with added bursts of star formation with varying duration and relative strength was used to model the star formation history of each galaxy.  Probability distribution functions were created for the relevant galaxy parameters after weighing each model by its $\exp(-\chi^{2}/2)$ and the median value for each of the parameters was used.  The SFRs have been averaged over the last 100 Myr and the $1\sigma$ uncertainties in stellar mass, when compared to both mock groups and other independent estimates, are on the order of 0.15 dex \citep{mcgee11}.  For the stellar masses probed in this work ($\log_{10}(M_{\rm{star}}/M_{\odot}) \geq 10$) there is no systematic offset between the \citet{bc03} (used for SDSS and GEEC) and \citet{bruzual07} (used for GEEC2) models.  The observed scatter between the two models is within our quoted $1\sigma$ uncertainties.  Additionally, there may be additional systematic uncertainties due to, for example, the IMF assumed in the fitting procedure. 

\subsection{Completeness Corrections}
\label{smlim}
\begin{figure*}
\centering
\includegraphics[width = 5.5cm, height = 5.5cm]{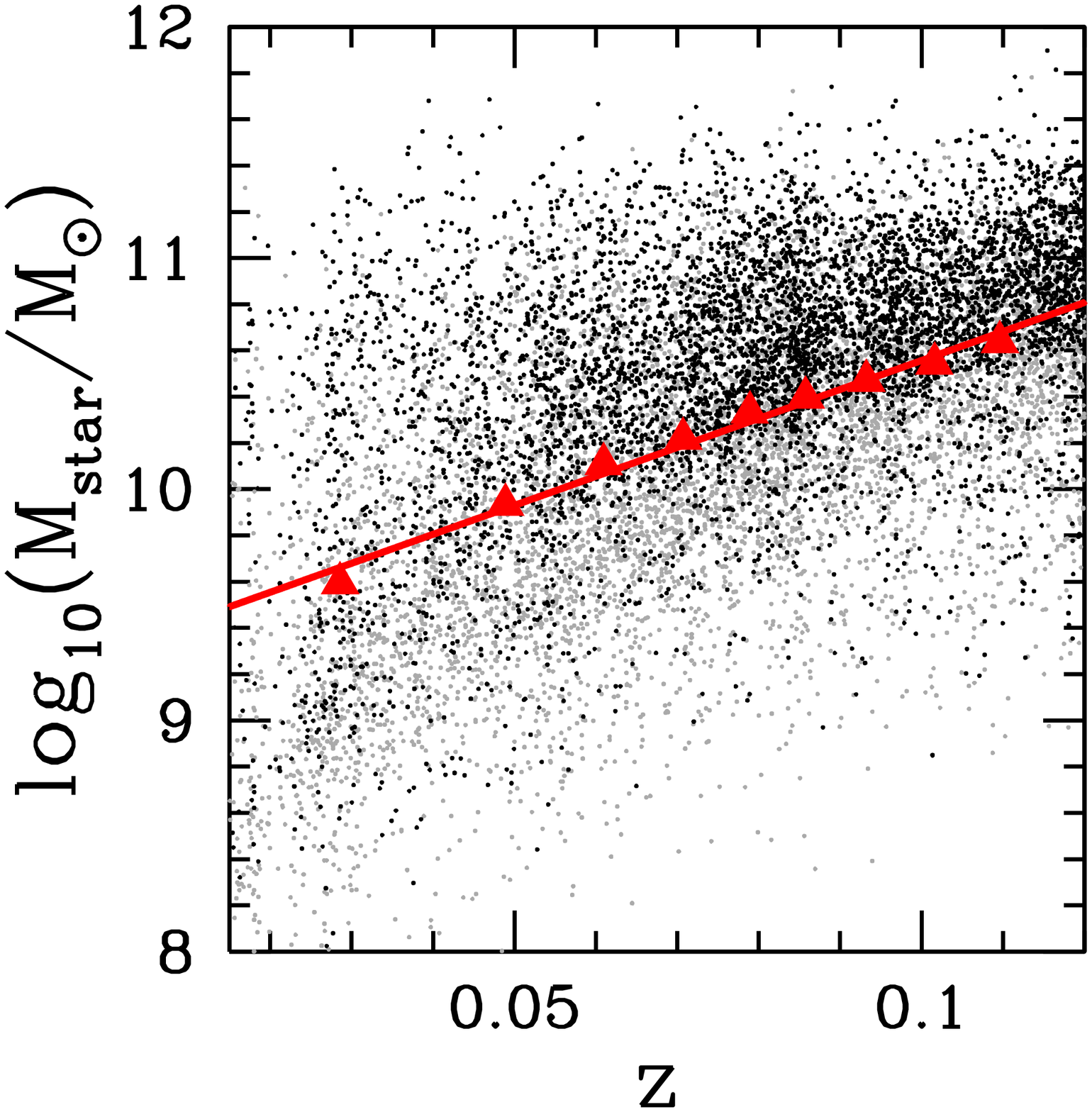}
\includegraphics[width = 5.5cm, height = 5.5cm]{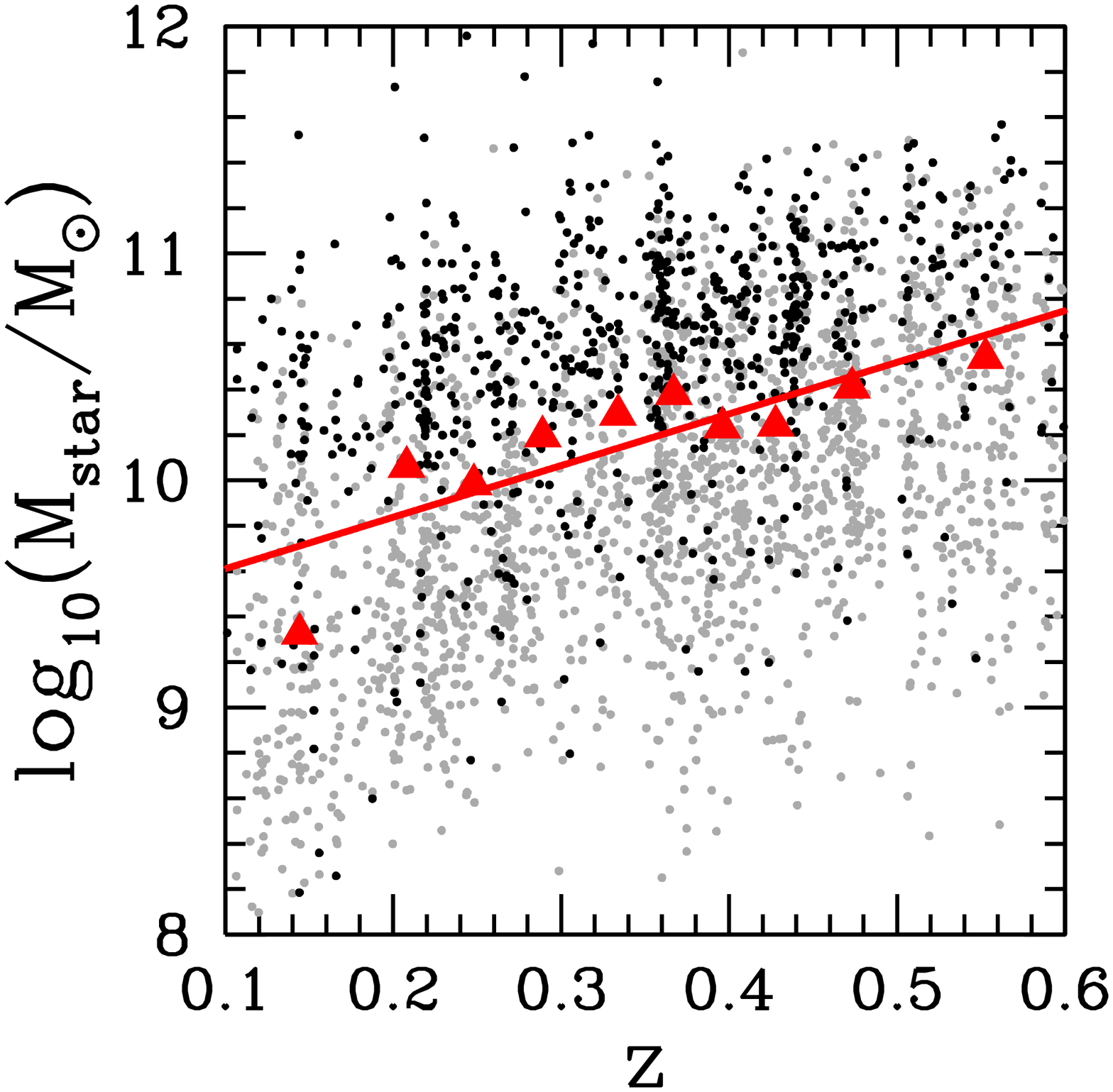}
\includegraphics[width = 5.5cm, height = 5.5cm]{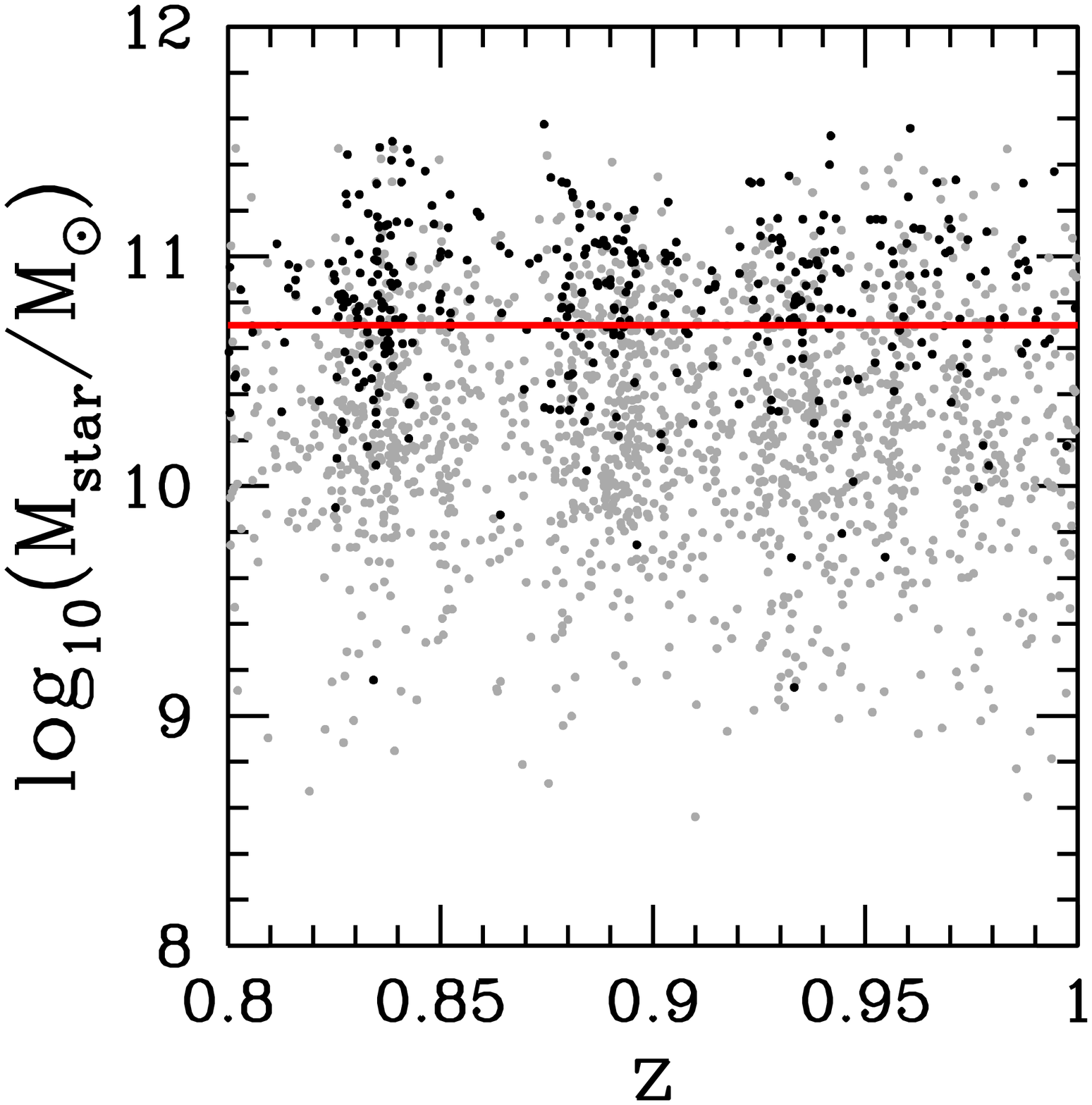}
\caption{Left: Observed $\log_{10}(M_{\rm{star}}/M_{\odot})$ versus $z$ for the quiescent galaxies (black circles) and for all galaxies (grey circles) in a sub-sample of 15 000 randomly selected galaxies in the SDSS catalogue.  The red triangles indicate the 90th percentile value of the stellar mass estimates of quiescent galaxies (black circles) given by Equation \ref{smcut} within a given redshift bin and the red solid line indicates a linear least-squares fit to these points.  This line is taken to be the stellar mass completeness limit of the sample.  Middle: Same as left except for all galaxies in the GEEC sample.  Right: Same as left except for all galaxies in the GEEC2 sample and the red solid line indicates a stellar mass cut of $\log_{10}(M_{\rm{star}}/M_{\odot}) = 10.7$. } 
\label{smcut}
\end{figure*}

The ability to detect faint and low mass galaxies declines with increasing redshift, which can be seen in Figure \ref{smcut} where we plot $\log_{10}(M_{\rm{star}}/M_{\odot})$ versus $z$ for all galaxies in each of the catalogues.  In order to address this stellar mass incompleteness, we apply a stellar mass limit to each of the group catalogues using the methodology described in \citet{connelly12}.   Briefly, we compute the stellar mass that each galaxy would have if it was observed at the $r$-band magnitude limit ($r_{\rm{lim}}$) of the sample using

\begin{equation}
\centering
M_{\text{star, \emph{r}$_{\rm{lim}}$}}(z) = M_{\rm{star}}(z)\times10^{\left(-0.4\left(r_{\rm{lim}} - r\left(z\right)\right)\right)},
\label{minmass}
\end{equation}
where $ M_{\rm{star}}(z)$ is the stellar mass of the galaxy determined from the SED fits and $r(z)$ is the observed $r$-band magnitude of the galaxy.

We define a conservative stellar mass limit by only taking the passive galaxies with SSFR $< 10^{-11}$ yr$^{-1}$ (SSFR $\equiv$ SFR/stellar mass)\footnote{It should be noted that \citet{connelly12} use red galaxies to define their limits}.  Since passive galaxies have on average a higher $M/L$ ratio than actively star-forming galaxies, we obtain a higher, and therefore more conservative, stellar mass limit using this methodology.  To define our limit, we compute the 90th percentile values of the mass estimates (Equation \ref{minmass}) for all passive galaxies in narrow redshift bins and then perform a linear least-squares fit to these values.  For the SDSS and GEEC catalogues, we then take all galaxies that fall above this line as our stellar mass complete catalogue.  To define our stellar mass completeness limit for the high redshift sample we take a different approach and apply a cut based on the GEEC2 sample selection criteria and the shape of the stellar mass function of the observed passive galaxy population \citep{mok13}.   For groups at the high redshift end ($z \sim 1$) of the GEEC2 catalogue, \citet{mok13} found that the sample was complete down to $\log_{10}(M_{\rm{star}}/M_{\odot}) = 10.7$ for passive galaxies.  We therefore take this value to be our stellar mass limit for the entire GEEC2 sample.  Although a limit of $\log_{10}(M_{\rm{star}}/M_{\odot}) = 10.7$ is conservative, we probe galaxy evolution via the quiescent fraction (see Section \ref{galprops}), which requires that the population of passive galaxies is complete.

In Figure \ref{smcut}, we plot the observed stellar masses versus redshift for passive galaxies (black circles) and all galaxies (grey circles) in the SDSS (left), GEEC (middle) and GEEC2 (right) surveys.  The 90th percentile stellar mass estimates for the SDSS and GEEC samples are shown as red triangles in Figure \ref{smcut} and the linear least-squares fit to these values as the red solid line.  For the GEEC2 sample (Figure \ref{smcut}: right), the red solid line indicates our stellar mass limit of $\log_{10}(M_{\rm{star}}/M_{\odot}) = 10.7$.  The stellar mass ranges for our complete samples are;  $9.5 \lesssim \log_{10}(M_{\rm{star}}/M_{\odot}) \lesssim 11.5$ for SDSS, $9.6 \lesssim \log_{10}(M_{\rm{star}}/M_{\odot}) \lesssim 11.5$ for GEEC and $10.7 \leq \log_{10}(M_{\rm{star}}/M_{\odot}) \lesssim 11.5$ for GEEC2.

In addition, we have applied a spectroscopic completeness correction by calculating magnitude weights for each galaxy.  The weights are computed following a methodology similar to \citet{wilman05}, where weights are calculated in $r$-band magnitude bins down to the limiting magnitude of each catalogue.

\subsection{Final group membership}
\label{finalgroups}
To probe the effects of groups dynamics on the properties of members galaxies, we only consider the sample of group galaxies with measured stellar masses and SFRs.  In addition, we only look at groups with more than five member galaxies within two virial radii ($r_{200}$) of the group centroid, where $r_{200}$ is defined as \citet{carlberg97}

\begin{equation}
\centering
r_{200} = \frac{\sqrt{3}\sigma_{\rm{rest}}}{10H(z)},
\label{r200}
\end{equation}
where $\sigma_{\rm{rest}}$ is the observed velocity dispersion ($\sigma_{\rm{obs}}$), computed via the Gapper Estimator \citep{beers90} from all member galaxies within 1.0 Mpc of the group centroid, corrected for redshift  (i.e.\ $\sigma_{\rm{rest}} = \sigma_{\rm{obs}}/(1 + z))$ and $H(z) = H_{0}\sqrt{\Omega_{m,0}(1 + z)^{3} + \Omega_{\Lambda,0}}$.

The inclusion of galaxies out to $2r_{200}$ is motivated by previous results.  In \citet{hou12}, we found that substructure galaxies were preferentially located on the group outskirts, beyond the virial radius.  Therefore, in order to better study correlations between the amount of substructure and galaxy properties, we include galaxies out to $2r_{200}$.  We discuss the effects of applying different radial cuts in Section \ref{results}.  

\begin{figure}
\centering
\includegraphics[width = 8cm, height = 8cm]{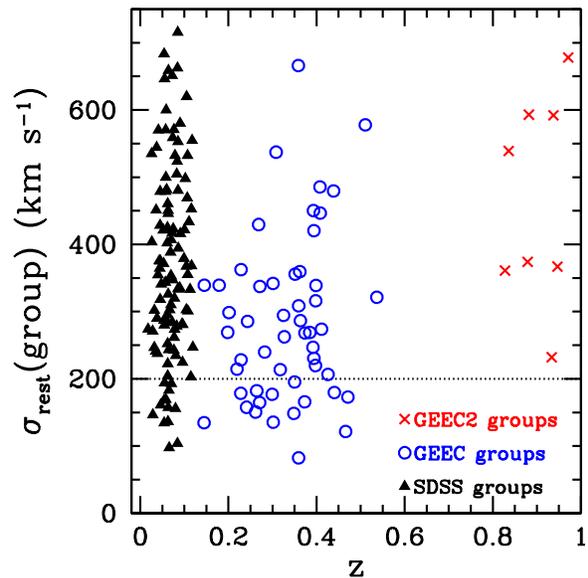}
\caption{Group velocity dispersion ($\sigma_{\rm{rest}}$) versus redshift ($z$) for our sub-sample of groups with $n_{\rm{members}} \geq 5$ within $2r_{200}$ of the group centroid identified in SDSS (black triangles), GEEC (blue circles) and GEEC2 (red crosses).  The dashed line indicates the lower limit $\sigma_{\rm{rest}}$ cut; we only analyze groups with $\sigma_{\rm{rest}} > 200$ km s$^{-1}$.}
\label{sigzall}
\end{figure}

In Figure \ref{sigzall} we plot the group velocity dispersion ($\sigma_{\rm{rest}}$) versus redshift ($z$) for our sub-sample of the three group catalogues.  The SDSS, GEEC and GEEC2 groups span a wide range of velocity dispersions, and therefore masses.  From Figure \ref{sigzall}, we see that both the SDSS and GEEC group catalogues contain lower mass systems when compared to GEEC2; therefore, to ensure that all three catalogues span a similar mass range we only consider groups with $\sigma_{\rm{rest}} > 200$ km s$^{-1}$, which corresponds to a dynamical mass of $\sim 1.2 \times 10^{13} M_{\odot}$ at a redshift of $z = 0.25$.  The minimum of five members within $2r_{200}$ requirement, and the $\sigma_{\rm{rest}} > 200$ km s$^{-1}$ cut, leaves us with 100 SDSS groups, 37 GEEC groups and 8 GEEC2 groups (see Table \ref{sedtab}).  

\subsection{Characterizing the properties of galaxies}
\label{galprops}
In order to study the relationship between environment and galaxy evolution, we look at the specific star formation rates of the galaxies in groups.  We examine both the SSFR distributions and the fraction of quiescent galaxies (hereafter $f_{\rm{q}}$), where $f_{\rm{q}}$ is defined as
\begin{equation}
\centering
f_{\rm{q}} = \frac{\text{$\#$ galaxies with SSFR} < 10^{-11} \text{yr}^{-1}}{\text{total $\#$ of galaxies}},
\label{fpass}
\end{equation}
with  SSFR $= 10^{-11}$ yr$^{-1}$ marking the division between the main sequence of star-forming galaxies from the quiescent galaxies in the SSFR-stellar mass plane \citep{mcgee11}.  It should also be noted that values in Equation \ref{fpass} are weighted to account for spectroscopic incompleteness.

\section{Galaxy properties with galaxy stellar mass and group dynamical mass}
\label{analysis}
\subsection{Correlations with galaxy stellar mass}
\label{nature}
It is well known that the observed properties of galaxies correlate well with galaxy stellar mass.  Many studies have shown that there exists a SFR-stellar mass trend, which is especially strong for star-forming galaxies \citep{kennicutt83, brinchmann04, noeske07, whitaker12}, and a colour-stellar mass trend \citep{tortora10, giodini12}, where massive galaxies are typically redder and have lower SFRs.  Before we investigate the role of group dynamics in galaxy evolution, we must first characterize the stellar mass trend in our sample. 
 
\begin{figure}
\centering
\includegraphics[width = 7.5cm, height = 7.5cm]{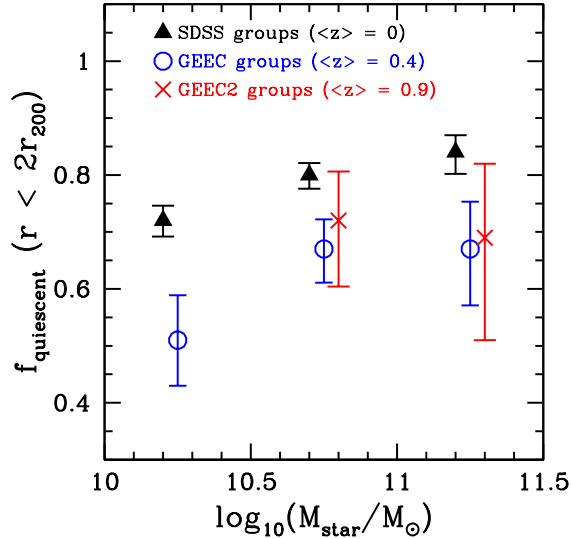}
\caption{Quiescent fraction ($f_{\rm{q}}$) versus stellar mass for all group galaxies in SDSS (black triangles), GEEC (blue circles) and GEEC2 (red crosses).  The data are divided into stellar mass bins of 0.5 dex in the range of $ 10 \leq \log_{10}(M_{\rm{star}}/M_{\odot}) \leq 11.5$ and plotted at the center of the mass range with small horizontal offsets for clarity.  It should be noted that due to our stellar mass cuts, the intermediate mass galaxies in the GEEC2 sample span a stellar mass range of $10.7 \leq \log_{10}(M_{\rm{star}}/M_{\odot}) < 11$.  Also, note that all catalogues are stellar mass complete and have been corrected for spectroscopic incompleteness.  The uncertainties in the quiescent fraction are computed following the methodology of \citet{cameron11}.}
\label{fqsmass}
\end{figure}

We look at the SSFR-stellar mass trend in each of our three group catalogues.  In Figure \ref{fqsmass} we plot $f_{\rm{q}}$ (Equation \ref{fpass}) versus stellar mass for all galaxy group members in SDSS (black triangles), GEEC (blue circles) and GEEC2 (red crosses).  From Figure \ref{fqsmass}, we see that the quiescent fraction shows a positive correlation with stellar mass for the SDSS and GEEC samples, as previously noted by \citet{mcgee11}.  For all three catalogues, the $f_{q}$-stellar mass trend appears to be flat for galaxies with $\log_{10}(M_{\rm{star}}/M_{\odot}) > 10.5$.  

An additional trend that can be seen in Figure \ref{fqsmass} is that for low mass galaxies ($\log_{10}(M_{\rm{star}}/M_{\odot}) < 10.5$) we observe an evolution in the quiescent fraction with redshift, where galaxies at higher redshifts have lower $f_{q}$.  We see a similar, though less drastic, trend for the massive galaxies ($\log_{10}(M_{\rm{star}}/M_{\odot}) > 10.5$) when comparing our $z \sim 0$ and $z \sim 0.4$ samples.  However, we do not observe a clear evolution in the quiescent fractions of massive galaxies between $z \sim 0.4$ and $z \sim 0.9$.  

In Section \ref{smremove}, we discuss how we remove this strong correlation between quiescent fraction and stellar mass so that we can examine the effects of group dynamics.   
 
\subsection{Correlations with group dynamical mass ($M_{200}$)}
\label{globalenv}
There are a number of possible processes related to environmentally driven galaxy evolution, including: ram-pressure stripping, strangulation and galaxy-galaxy interactions.  Some of these mechanisms are more directly related to the potential of the group, while others are better correlated to the local or neighbouring environment.  To probe the influence of the host group, we look for correlations between the observed quiescent fraction and dynamical mass, $M_{200}$, of the group defined as \citet{carlberg97}
 
\begin{equation}
\centering
M_{200} = \frac{3r_{200}\sigma_{\rm{rest}}^{2}}{G},
\label{m200}
\end{equation}
where $r_{200}$ is given by Equation \ref{r200}.  It should be noted that the mass computed in Equation \ref{m200} assumes that the system is in dynamical equilibrium, which we show in Section \ref{grpdyn} is not always true for the groups in our catalogues.  \citet{bird95} showed that dynamical mass estimators, such as Equation \ref{m200}, tend to overestimate the true mass of systems not in equilibrium, in particular those with significant substructure.  However, our goal is to roughly divide our sample by mass and this methodology works well for this purpose.  Alternatively, we could have used the total stellar mass of the group to characterize the host environment, though this method requires significant completeness corrections.  It should be noted that we do observe similar results whether $M_{200}$ or total stellar mass is used in the analysis.  

\begin{figure}
\centering
\includegraphics[width = 8.7cm, height = 8.7cm]{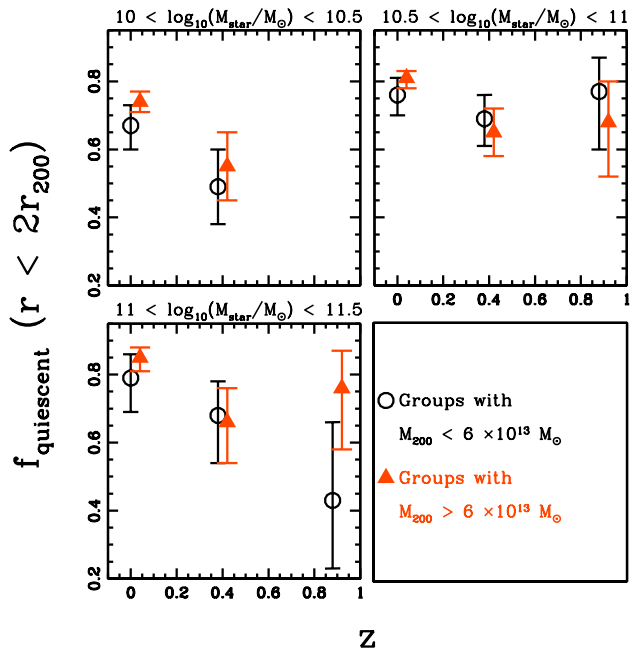}
\caption{$f_{\rm{q}}$ versus $z$ for low mass galaxies ($10 < \log_{10}(M_{\rm{star}}/M_{\odot}) \leq 10.5$: top left), intermediate mass galaxies ($10.5 < \log_{10}(M_{\rm{star}}/M_{\odot}) \leq 11$: top right) and high mass galaxies ($11 < \log_{10}(M_{\rm{star}}/M_{\odot}) \leq 11.5$: bottom left) in low mass ($M_{200} < 6 \times 10^{13} M_{\odot}$) groups (black circles) and in high mass ($M_{200} > 6 \times 10^{13} M_{\odot}$) groups (orange triangles).  It should be noted that due to our stellar mass cuts, the intermediate mass galaxies in the GEEC2 sample span a stellar mass range of $10.7 \leq \log_{10}(M_{\rm{star}}/M_{\odot}) < 11$.  We plot the data at the redshift each sample has been $k$-corrected to: $z = 0$ for SDSS, $z = 0.4$ for GEEC and $z = 0.9$ for GEEC2, with small horizontal offsets for clarity.  We remind the reader of the redshift range for each catalogue: $0 < z < 0.12$ for SDSS, $0.15 < z < 0.55$ for GEEC and $0.8 < z < 1$ for GEEC2.  The uncertainties in quiescent fraction are computed following the methodology of \citet{cameron11}.}
\label{dynmassbins}
\end{figure}

We make a cut at $M_{200} = 6 \times 10^{13} M_{\odot}$ to distinguish between low and high mass groups as this is the approximate median value for each of the three group catalogues.  In Figure \ref{dynmassbins}, we plot $f_{\rm{q}}$ versus $z$ for  low mass galaxies ($10 < \log_{10}(M_{\rm{star}}/M_{\odot}) \leq 10.5$: top left), intermediate mass galaxies ($10.5 < \log_{10}(M_{\rm{star}}/M_{\odot}) \leq 11$: top right) and high mass galaxies ($11 < \log_{10}(M_{\rm{star}}/M_{\odot}) \leq 11.5$: bottom left) in low mass ($M_{200} < 6 \times 10^{13} M_{\odot}$) groups (black circles) and in high mass ($M_{200} > 6 \times 10^{13} M_{\odot}$) groups (orange triangles).  From Figure \ref{dynmassbins}, we see that for all stellar masses the quiescent fraction of galaxies in low and high mass groups are not statistically distinct.  It should be noted that if we make an additional cut at $M_{200} = 10^{14} M_{\odot}$, we still find no dependence of $f_{q}$ on group halo mass.  While numerous studies have found that the observed properties of galaxies do correlate with halo mass \citep{pasquali10, wetzel12}, these studies also show that the trends tend to be flatter for higher mass haloes.  In particular, given the dynamical mass range of the SDSS, GEEC and GEEC2 groups ($13 \lesssim \log_{10}(M_{200}/M_{\odot}) \lesssim 14.5$) and our average errors on $f_{q}$ of $\sim \pm 10$ per cent, the average quiescent fractions and ages shown in \citet{wetzel12} and \citet{pasquali10} are approximately the same for galaxies in low and high mass groups, assuming a cut at  $M_{200} = 6 \times 10^{13} M_{\odot}$.  

\section{Group Dynamics}
\label{grpdyn}
Having considered how the observed quiescent fraction of galaxies in groups correlates with galaxy stellar mass and group dynamical mass, $M_{200}$, we now examine how the dynamical state of a group affects the properties of member galaxies.  Previous studies of the local environment have been characterized by the local density \citep[e.g.,][]{poggianti08} and by the number of nearest neighbours, typically 3-10 \citep[e.g.,][]{gomez03, patel11, sobral11}.  While these methods are effective in determining the local over-density of regions within groups, they are not directly related to the dynamical state of a group.  With a spectroscopic group catalogue we can directly measure the dynamical state of the group, both in terms of the local environment and the host group halo.  In the following section, we describe how we classify the dynamical state of galaxy groups and present our analysis of the SDSS, GEEC and GEEC2 groups.

\subsection{Determining the dynamical state of groups}
\label{tests}
We classify the dynamical state of a group using two methods:

\begin{enumerate}
\item The shape of the group velocity distribution;
\item The amount of substructure.
\end{enumerate}
Theoretically, a system in dynamical equilibrium (i.e.\ relaxed or virialized) should have a Gaussian velocity distribution; thus, deviations from such a distribution would indicate a dynamically complex or unevolved system.  In \citet{hou09}, we showed that we can reliably and robustly identify non-Gaussian velocity distributions for systems with as few as five member galaxies using the Anderson-Darling (AD) goodness-of-fit test.  A full description of the AD Test, and its application to group-sized systems, is given in \citet{hou09}.  For our analysis, we use the AD statistic to compare the cumulative distribution function (CDF) of the ordered galaxy velocities to a Gaussian distribution using the computing formula given by

\begin{equation}
\centering
A^{2} = -n - \frac{1}{n}\sum_{i = 1}^{n}\left(2i - 1\right)\left(\ln\Phi\left(x_{i} \right)+ \ln\left(1 - \Phi\left(x_{n + 1  - i}\right)\right)\right),
\end{equation}
\begin{equation}
\centering
A^{2\ast} = A^{2}\left(1 + \frac{0.75}{n} + \frac{2.25}{n^{2}}\right),
\end{equation}
where $x_{i} \leq x \leq x_{i + 1}$, $\Phi(x_{i})$ is the CDF of a Gaussian distribution \citep{d'agostino}.  The probabilities, or $p$-values, are then computed as 
\begin{equation}
\centering
p = a \exp(-A^{2\ast}/b),
\end{equation}
where a = 3.6789468 and b = 0.1749916, and both factors are determined via Monte Carlo methods \citep{nelson98}.  We then classify groups as dynamically complex if the group velocity distribution is identified as non-Gaussian at the 95 per cent confidence level ($p$-value $< 0.05$).

We also examine the amount of substructure present in each group by applying the Dressler-Shectman (DS) Test \citep{ds88} to our group samples.  Substructure is indicative of the recent accretion of galaxies or smaller groups of galaxies \citep{lc93}.  In \citet{hou12}, we showed that the DS test, originally developed for richer galaxy clusters, could robustly identify substructure for groups with $n_{\rm{members}} \geq 20$.  Additionally, we found that the test could be applied to groups with as few as 10 members, but in this case the measured fraction of systems with substructure is underestimated.  A detailed description of the test, with respect to group-sized systems can be found in \citet{pinkney96}, \citet{zm98a} and \citet{hou12}.  Briefly, for each group we compute the mean velocity ($\overline{\nu}$) and group velocity dispersion ($\sigma$).  Then, for each member galaxy $i$, we compute the local mean velocity ($\overline{\nu_{local}^{i}}$) and local velocity dispersion ($\sigma_{local}^{i}$) using the $i$th galaxy plus a number of its nearest neighbours ($N_{nn}$).  Using these values we then compute 

\begin{equation}
\centering
\delta_{i} = \left(\frac{N_{nn} + 1}{\sigma^{2}}\right)\left[\left(\overline{\nu_{local}^{i}} - \overline{\nu}\right)^{2} + \left(\sigma_{local}^{i} - \sigma\right)^{2}\right],
\end{equation}
where $1 \leq i \leq n_{\rm{members}}$ and $N_{nn} = \sqrt{n_{\rm{members}}}$, rounded down to the nearest integer.  The DS statistic is then computed as
\begin{equation}
\centering
\Delta = \sum_{i = 1}^{n}\delta_{i}.
\label{DS}
\end{equation}
We use Monte Carlo methods to determine the probability or $p$-value for the DS Test, which is done by comparing the observed $\Delta$-value to `shuffled $\Delta$-values', which are computed by randomly shuffling the observed velocities and then re-assigning them to the observed member galaxy positions.  The $p$-value is then calculated as 
\begin{equation}
\centering
p = \sum (\Delta_{\text{shuffled}} > \Delta_{\text{observed}})/n_{\text{shuffle}},
\end{equation}
where $\Delta_{\text{shuffled}}$ and $\Delta_{\text{observed}}$ are both computed with Equation \ref{DS}.  We compute the $p$-value using 100 000 shuffled $\Delta$-values.  A group is identified as having significant substructure if it has a $p$-value $<$ 0.05.

Following this methodology, we classify the dynamical state of the SDSS, GEEC and GEEC2 groups using the AD Test for all groups with $n_{\rm{members}} \geq 5$ and the DS Test for $n_{\rm{members}} \geq 10$.  In Table \ref{dynevo}, we list the results of the tests, where we find the percentage of non-Gaussian groups in the SDSS, GEEC and GEEC2 surveys to be 15 $\pm^{6}_{4}$ per cent, 51 $\pm 11$ per cent and 25 $\pm^{26}_{13}$ per cent, while the percentage of groups with detected substructure remains approximately constant at $\sim20$ per cent for all three group catalogues.  

For completeness, we include the AD and DS Test results for the GEE2 sample, however it should be noted that with such a small sample of systems we cannot robustly determine the fraction of non-Gaussian groups and groups with substructure.

\begin{table}
\centering
\caption{Fraction of dynamically complex (non-Gaussian) groups and groups with substructure using the AD and DS Tests. \label{dynevo}}
\begin{tabular}{ccc}
\hline\hline
Catalogue & Fraction of & Fraction of Groups\\
       & non-Gaussian groups & with substructure\\
\hline
SDSS & 15/100 (15 $\pm^{6}_{4} \%$) & 17/71 (24 $\pm^{8}_{6} \%$) \\
GEEC & 19/37 (51$\pm 11 \%$) & 3/14 (21 $\pm^{19}_{10} \%$)\\
GEEC2 & 2/8 (25 $\pm^{26}_{13} \%$) & 1/5 (20 $\pm^{34}_{12} \%$)\\
\hline
\end{tabular}
\end{table}

\subsection{Dynamics and galaxy properties}
\label{results}
Having classified the dynamical state of the groups in our sample, we now compare the SSFR distributions and quiescent fractions of the galaxies in groups that are dynamically young to those in dynamically evolved systems.  

\subsubsection{SSFR Distributions}
\begin{table*}
\centering
\caption{Probabilities ($p$-values) from a two-sample KS Test comparing the SSFR distributions shown in Figure \ref{ssfrs}.   Probabilities $<$ 0.01 indicate that the systems come from different underlying parent distributions.  \label{ssfrks}}
\begin{tabular}{ccc}
\hline\hline
Catalogue & $p$-value comparing Gaussian & $p$-value comparing groups with \\
    & versus non-Gaussian groups & substructure vs. no substructure\\
\hline
SDSS & 0.01099 & 0.4729\\
GEEC & $\sim 0$ & $\sim 0$\\
GEEC2 & 0.6869  & \\
\hline
\end{tabular}
\end{table*}
\label{globalfrac}

\begin{figure*}
\centering
\includegraphics[scale= 0.6]{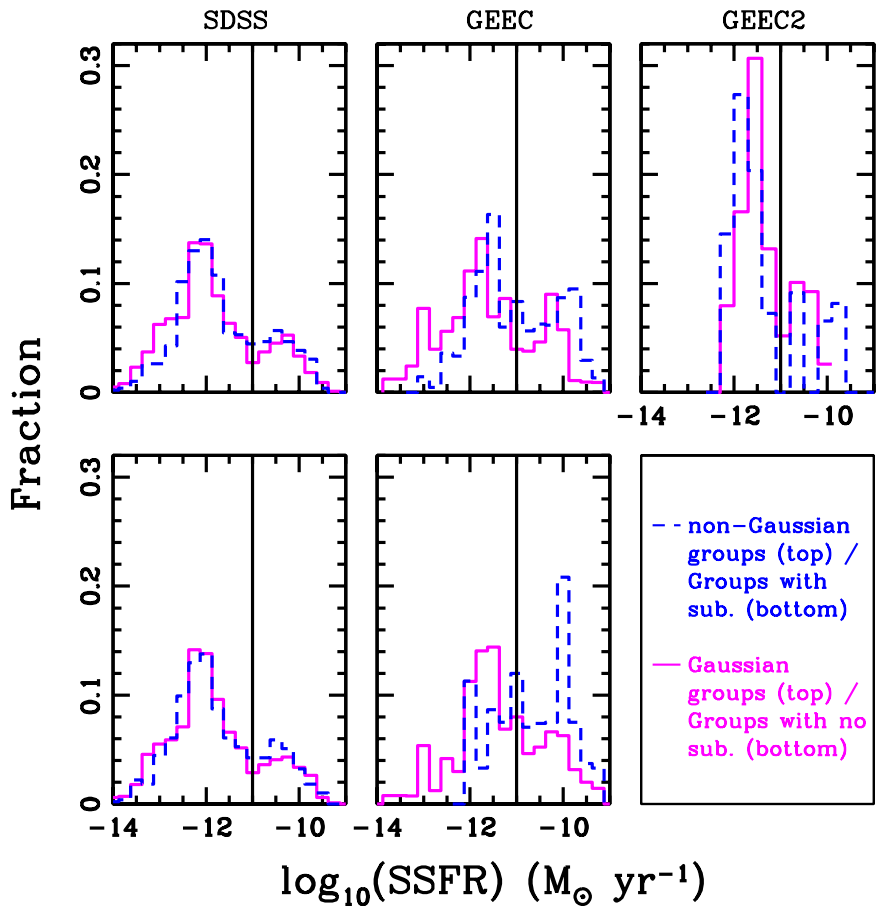}
\caption{Top: SSFR distributions for galaxies in non-Gaussian groups (blue dashed line) and for galaxies in Gaussian groups (magenta solid line) in the SDSS catalogue (left), GEEC catalogue (middle) and GEEC2 catalogue (right).  Note that all catalogues are stellar mass complete and spectroscopic completeness weights have been taken into account.  Bottom:  SSFR distributions for galaxies in groups with substructure (blue dashed line) and for galaxies in groups with no identified substructure (magenta solid line) for the SDSS (left) and GEEC (middle) sample.  We do not show the SSFR distributions for the GEEC2 groups with and without substructure, as our stellar mass limit and $n \geq 10$ within $2r_{200}$ cut for the DS Test result in too few galaxies.}
\label{ssfrs}
\end{figure*}
In the top panels of Figure \ref{ssfrs} we plot the SSFR distributions for galaxies in non-Gaussian (blue dashed line) and Gaussian (solid magenta line) groups identified in the SDSS (left), GEEC (middle) and GEEC2 (right) group catalogues.  The bottom panels of Figure \ref{ssfrs} are the same except we plot the SSFR distributions for galaxies in groups with substructure (blue dashed line) and galaxies in groups with no substructure (solid magenta line) for the SDSS (left) and GEEC (middle) samples.  We do not show the SSFR distributions for the GEEC2 groups with and without substructure because the sample contains too few galaxies.  For the same reason we do not include the GEEC2 groups with and without substructure in our analysis for the remainder of the paper.

In all panels of Figure \ref{ssfrs}, we see that the histograms are bimodal, showing a population of actively star-forming galaxies with  SSFR $> 10^{-11}$ yr$^{-1}$ and a population of passive galaxies with  SSFR $< 10^{-11}$ yr$^{-1}$.  For the SDSS groups, it appears that the SSFR distributions for dynamically complex and relaxed systems, classified with both the AD- and DS-Test, are similar with a well populated passive sequence.  However, a two-sample Kolmogorov-Smirnov (KS) Test indicates that while the SSFR distributions for the SDSS galaxies in groups with and without substructure likely come from the same parent distribution, the distributions for galaxies in Gaussian and non-Gaussian groups are in fact distinct at the $\sim$99 per cent confidence level (Table \ref{ssfrks}).  Though it should be noted that difference is small and it is easier to detect small differences given the large size of the SDSS sample.  

Looking at the $z \sim 0.4$ GEEC groups (middle panels of Figure \ref{ssfrs}), we see that the SSFR distributions for the dynamically complex and relaxed groups, identified with either the AD- or DS-Test, look distinct.  Indeed, a two-sample KS Test indicates that both sets of SSFR distributions come from different parent distributions at a confidence level of $>$99 per cent (Table \ref{ssfrks}).  For the Gaussian and non-Gaussian groups, we see that although both histograms show a bimodal distribution with a well populated passive sequence, there are more galaxies with high SSFR's ($\sim 10^{-10}$ yr$^{-1}$: Figure \ref{ssfrs}) in the non-Gaussian GEEC groups.  The GEEC groups with no detected substructure show a well populated passive sequence, while the majority of galaxies in the GEEC groups with substructure appear to lie in the actively star-forming sequence.  A similar result was shown in \citet{hou12}, where we found that galaxies in groups with substructure had a significantly higher fraction of blue galaxies.

The SSFR distributions for the galaxies in Gaussian and non-Gaussian GEEC2 groups are consistent with coming from the same parent distribution (Table \ref{ssfrks}) and show similar features to the SDSS groups (i.e.\ a dominant quiescent population).  The high fraction of quiescent galaxies is likely a result of our stellar mass completeness limits  ($\log_{10}(M_{\rm{star}}/M_{\odot}) \geq 10.7$), which from Figure \ref{fqsmass} would result in a more dominant quiescent population.

\subsubsection{Quiescent Fractions}
\label{smremove}
\begin{figure*}
\centering
\includegraphics[width = 8.7cm, height = 8.7cm]{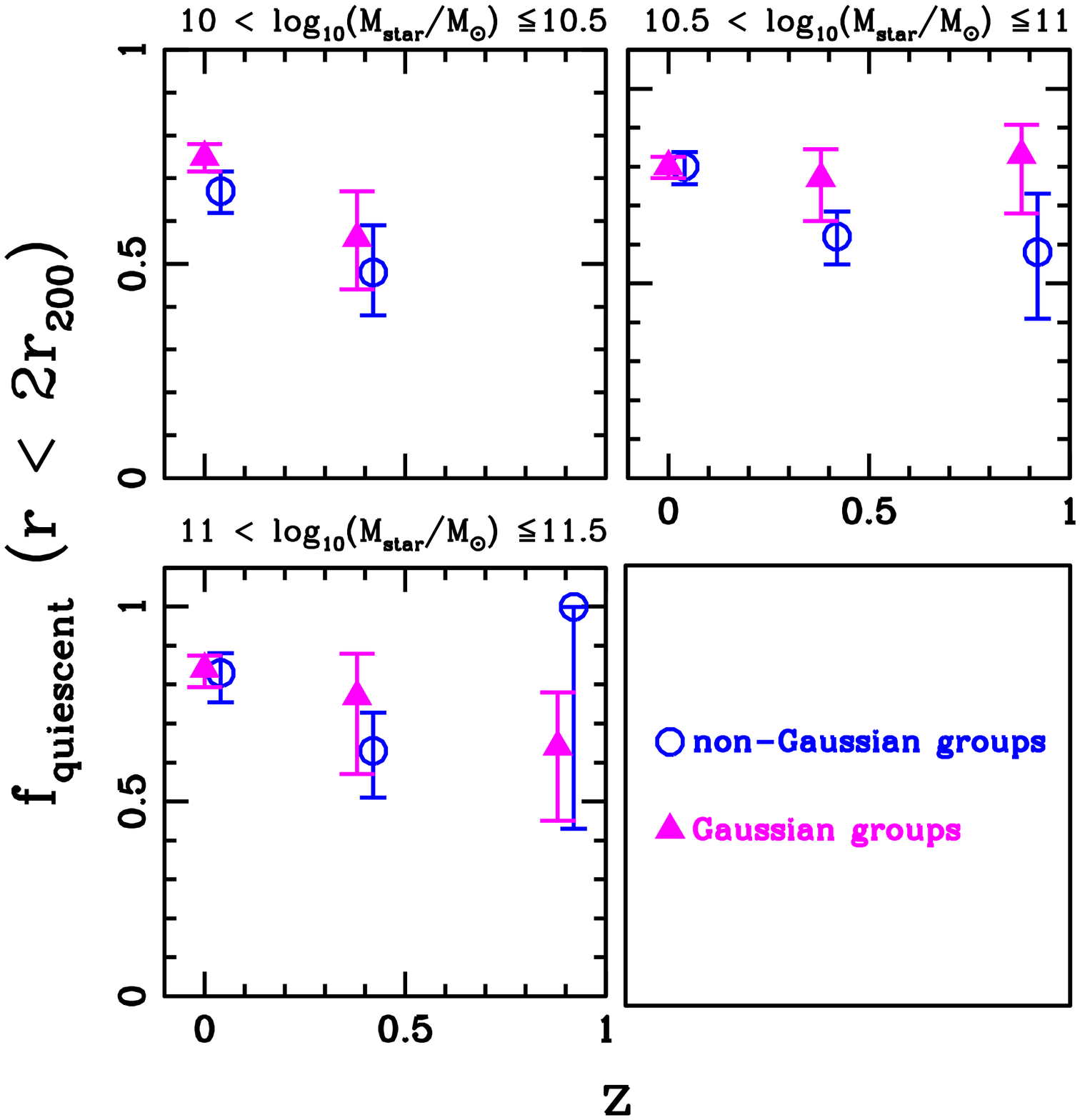}
\includegraphics[width = 8.7cm, height = 8.7cm]{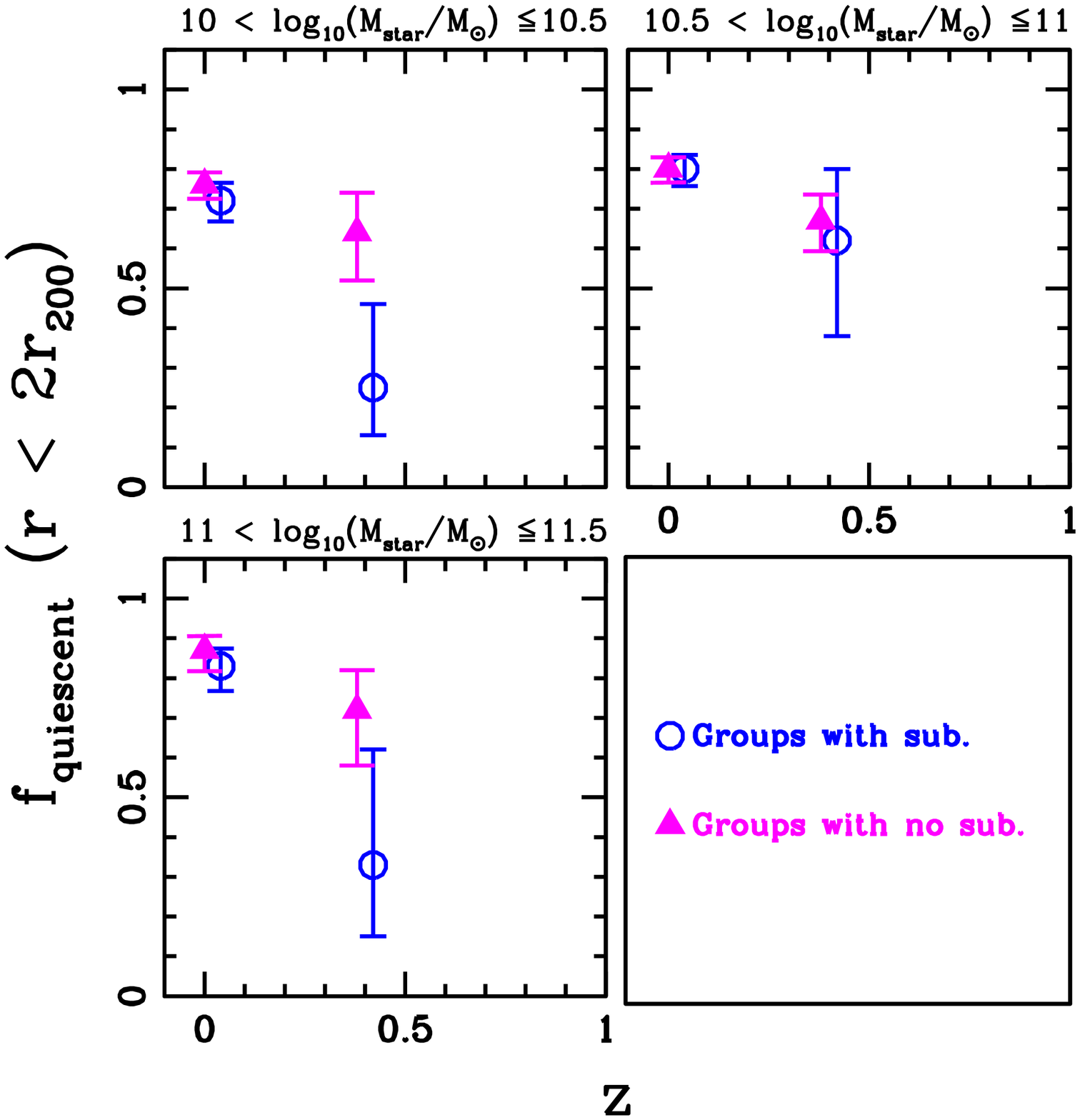}
\caption{Left:  $f_{\rm{q}}$ versus $z$ for galaxies in groups with non-Gaussian velocity distributions (blue circles) and in groups with Gaussian velocity distributions (magenta triangles).  The panels are divided into bins of stellar mass, with; $10 < \log_{10}(M_{\rm{star}}/M_{\odot}) \leq 10.5$ (top-left), $10.5 < \log_{10}(M_{\rm{star}}/M_{\odot}) \leq 11$ (top-right) and $11 < \log_{10}(M_{\rm{star}}/M_{\odot}) \leq 11.5$ (bottom-left).  Note that all catalogues are spectroscopic and stellar mass complete.  Also, due to the stellar mass limits the intermediate mass bin for the GEEC2 sample does not extend down to $\log_{10}(M_{\rm{star}}/M_{\odot}) \leq 10.5$, but rather covers a range of $10.7 \leq \log_{10}(M_{\rm{star}}/M_{\odot}) \leq 11$.  We plot the data at the redshift that each sample has been $k$-corrected to: $z = 0$ for SDSS, $z = 0.4$ for GEEC and $z = 0.9$ for GEEC2, with small horizontal offsets for clarity.  Right:  Same as figure on the left except for galaxies in groups with substructure (blue circles) and in groups with no significant substructure (magenta triangles).  Again, we do not show the GEEC2 groups with and without substructure, as our stellar mass limits and $n \geq 10$ within $2r_{200}$ cut for the DS Test resulted in too few galaxies in each sub-sample.  The uncertainties in quiescent fraction are computed following the methodology of \citet{cameron11}.}
\label{smassbins}
\end{figure*}

We now look at the quiescent fraction ($f_{q}$: Equation \ref{fpass}) of galaxies in dynamically complex and relaxed groups.  In the left panels of Figure \ref{smassbins}, we plot $f_{\rm{q}}$ versus $z$ for low mass galaxies ($10 < \log_{10}(M_{\rm{star}}/M_{\odot}) \leq 10.5$: top left), intermediate mass galaxies ($10.5 < \log_{10}(M_{\rm{star}}/M_{\odot}) \leq 11$: top right) and high mass galaxies ($11 < \log_{10}(M_{\rm{star}}/M_{\odot}) \leq 11.5$: bottom left) in non-Gaussian groups (blue symbols) and in Gaussian groups (magenta triangles).  The panels on the right are similar except we plot galaxies in groups with substructure (blue symbols) and in groups with no significant substructure (magenta triangles) but only for the SDSS and GEEC groups.  The GEEC2 groups are omitted as there are too few galaxies for a robust substructure analysis.  In order to isolate the effects of dynamical state on the properties of galaxies from the strong $f_{\rm{q}}$-stellar mass trend (Figure \ref{fqsmass}), we bin our data into narrow bins of stellar mass.

Looking at Figure \ref{smassbins}, we find that at almost all epochs and stellar masses there is no significant difference in the quiescent fractions of galaxies in dynamically complex and relaxed groups for both dynamical classification schemes (AD and DS Tests).  However, we do observe a difference in the low mass bin ($10 < \log_{10}(M_{\rm{star}}/M_{\odot}) \leq 10.5$) of the GEEC sample, where the groups with substructure have a lower $f_{q}$ than observed in the groups with no substructure.  

It should also be noted that our analysis was done using galaxies within two viral radii of the group centroid; however, if we use only galaxies with $r < r_{200}$ we find similar results.  Although including galaxies beyond the virial radius ($r_{200}$) inherently means that we are investigating the `unvirialized' regions of our systems, we find that while the fraction of dynamically young systems increases within each sample, the trends with redshift remain the same whether we use $r_{200}$ or $2r_{200}$.  In \citet{hou12}, we determined that substructure galaxies were preferentially found on the group outskirts.  Thus, analyzing galaxies out to two virial radii allows us to better study substructure in our groups.  Additionally, studies have shown that the effects of the environment on galaxies can extend well beyond the virial radius \citep{feldmann10, vdl10,bahe12}.

\section{Discussion}
\label{discussion}
In Section \ref{grpdyn} we classified the dynamical state of our group sample and then compared the SSFR distributions and quiescent fractions of galaxies in dynamically complex and relaxed groups.  We now discuss the implications of our findings.   

\subsection{The evolution of group dynamics}
In a $\Lambda$CDM Universe structure grows hierarchically through mergers and accretion \citep[e.g.,][]{springel05}.  Numerous studies have shown that at a given halo mass the average accretion rate of dark matter haloes goes as \emph{\.{M}/M} $\propto (1 + z)^{n}$, where $n \sim 1.5-2.5$ \citep{bdn07, mfm09}, indicating that the accretion rate increases with redshift.  As a reflection of this assembly history, one might expect the fraction of dynamically unevolved systems to increase with redshift for a given mass.  Although, additional factors, such as the time since infall or the mass and orbit of the accreted object, should also affect the evolution of the dynamical state.  For example, continuous accretion of smaller sub-haloes could result in less obvious deviations from a relaxed state in comparison to an instantaneous major merger of larger haloes \citep{cw05}.  Therefore, observations that are sensitive to different forms of mass assembly may result in different dynamical evolution scenarios.  However, based on a statistical study of a large sample of simulated $N$-body groups and clusters, identified with a FoF-algorithm, \cite{cw05} found that on average the virial ratio, 2KE/PE, of all clusters with $M > 10^{14} h^{-1}M_{\odot}$ increased with increasing redshift.  Therefore, systems at higher redshifts are more dynamically young or complex than in the local Universe.  Assuming that galaxies are a good tracer of the dark matter haloes, it should be possible to detect this predicted increase in dynamically unevolved systems.  

In Section \ref{tests}, we found that the fraction of groups with non-Gaussian velocity distributions, classified as dynamically young, increases significantly from $\sim$15 per cent at $z \sim 0$ to $\sim$51 per cent at $z \sim 0.4$ (Table \ref{dynevo}), which is in agreement with the results of \citet{cw05} who found that the virial ratio increased with redshift.  From $z \sim 0.4$ to $z \sim 0.9$ it appears that the fraction of non-Gaussian groups is consistent with either being flat or decreasing with increasing redshift (Table \ref{dynevo}); however, this result is based a small sample of 8 high-$z$ groups.  It is also important to note that the GEEC2 catalogue is different from the SDSS or GEEC samples in that: the groups were selected using a different methodology, all of the GEEC2 groups are X-ray bright while only some of the SDSS and GEEC groups are X-ray bright, and a different stellar mass completeness limit was applied.  Thus, the results of the GEEC2 sample may be due to small number statistics or differences in the sample selection.  Further investigation of a larger sample of high redshift groups is required to make any conclusive statements about the evolution of group dynamics from intermediate to high redshifts.  

We now look at the evolution of substructure in groups and we find that the fraction of groups with substructure is consistent out to $z \sim 1$.  These results appear to be contradictory.  However, the AD and DS Tests, though both measures of dynamical state, probe different stages of dynamical complexity \citep{pinkney96} and a 1-to-1 correspondence between non-Gaussian groups (identified from the AD Test) and groups with substructure (identified from the DS Test) does not necessarily hold.  In \citet{hou12} we showed that groups with substructure that is loosely bound or spatially mixed with members of the host group can be difficult to detect.  Therefore, groups with non-Gaussian profiles may have substructure that is missed by the DS test.  Also, since we studied groups with as few as 10 members, the results of the DS Test can only provide a lower limit on the fraction of groups with substructure \citep{hou12}, so the true fraction of groups with substructure is likely higher than the values quoted in Table \ref{dynevo}.  

\subsection{The effects of dynamics on galaxy properties}
We first look at the quiescent fraction as a function of redshift.  From Figures \ref{fqsmass} and \ref{smassbins} we see that for groups only the low mass galaxies ($10 < \log_{10}(M_{\rm{star}}/M_{\odot}) \leq 10.5$) clearly exhibit the well known Butcher-Oemler effect \citep{bo84, poggianti99, wilman05b, urquhart10,mcgee11, li12}, where $f_{q}$ decreases with increasing redshift.  In contrast, the quiescent fraction of intermediate and high mass galaxies ($10.5 < \log_{10}(M_{\rm{star}}/M_{\odot}) \leq 11.5$) in groups shows a marginal decrease between $z \sim 0$ to $z \sim 0.4$ and no obvious change between $z \sim 0.4$ to $z \sim 0.9$.  This result is similar to those of \citet{ra12}, who observed no increase in the fraction of high mass ($\log_{10}(M_{\rm{star}}/M_{\odot}) \gtrsim 11.13$) blue galaxies in clusters in the redshift range of $0 < z < 2.2$.  Based on our results, we find no clear evidence for the Butcher-Oemler effect in galaxies with $\log_{10}(M_{\rm{star}}/M_{\odot}) \gtrsim 10.5$ in groups out to $z \sim 1$.  However, we do observe decrease in the fraction of low mass ($\log_{10}(M_{\rm{star}}/M_{\odot}) < 10.5$) quiescent galaxies with increasing redshift. 

Finally, we examine the effects of dynamical state.  In general, we find that there is no correlation between dynamical state and quiescent fraction for massive galaxies ($\log_{10}(M_{\rm{star}}/M_{\odot}) > 10.5$); however, there may be a hint of a correlation with the presence of substructure in our lowest mass galaxies ($10 < \log_{10}(M_{\rm{star}}/M_{\odot}) \leq 10.5$ - Fig. \ref{smassbins}: right).  In our intermediate redshift GEEC sample, the groups with substructure have a lower quiescent fraction in comparison to galaxies in groups with no substructure.  Our results for the SDSS sample can be compared to the Zurich Environmental Study (ZENS) sample of \citet{carollo12} and are in good agreement.  \citet{carollo12} found that central galaxies and satelites with $\log_{10}(M_{\rm{star}}/M_{\odot}) > 10$ in dynamically relaxed and unrelaxed groups, classified via the DS Test, have similar observed galaxy properties.  However, these authors did find that satellites with $\log_{10}(M_{\rm{star}}/M_{\odot}) < 10$ are bluer by $\sim 0.1 \text{mag}$ in unrelaxed groups.  Our sample does not extend to these low masses, though we do find a similar result for slightly higer mass galaxies ($10 < \log_{10}(M_{\rm{star}}/M_{\odot}) \leq 10.5$) at intermediate redshifts ($z \sim 0.4$), which could indicate possible redshift evolution in the relationship between substructure and quiescent fraction.

In addition, several studies have also found that environmental effects on galaxy properties can only be observed in low mass galaxies \citep[$ \log_{10}(M_{\rm{star}}/M_{\odot}) \lesssim 10.5$ - ][]{peng10, sobral11}.  In particular, \citet{peng10} suggest that for low mass galaxies at $z \gtrsim 0.5$, the main mechanism responsible for star formation quenching is galaxy-galaxy interactions, which should be the dominant process in dynamically unevolved systems with significant substructure.  In addition, \citet{bb07} found that while the properties of star-forming galaxies were largely independent of environment, they did observe a correlation between colour and clustering on small scales ($< 300 h^{-1}$ Mpc), which they claim indicated that substructure within groups may play a role in the evolution of galaxies.  Similarly, \citet{wilman10} observed a correlation between the mean colour of blue galaxies and local density but only on the $\lesssim$ 1 Mpc scales, further suggesting that the local, and not global or large-scale, environment may have a more dominant affect on galaxy evolution. 

Although we do observe a difference in the quiescent fractions of low mass galaxies in GEEC groups with and without substructure, we note that this result is based on a small sample of groups (Table \ref{dynevo}).  A larger sample of intermediate and high redshift groups is required to make a more robust statement about whether quenching in low mass group galaxies is suppressed in the presence of substructure. 

\section{Conclusions}
\label{conclusion}
We have looked at the role of galaxy stellar mass, group dynamical mass ($M_{200}$) and dynamical state in the evolution of galaxies in groups out to $z \sim 1$ using the SDSS, GEEC and GEEC2 group catalogues.  The dynamical state of the groups are classified with the Anderson-Darling Test to distinguish between Gaussian and non-Gaussian velocity distributions and the Dressler-Shectman Test to determine the amount of substructure within the groups.  The main results of this analysis are:

\begin{enumerate}
\item We observe a strong trend between the quiescent fraction and galaxy stellar mass in SDSS and GEEC, where higher mass galaxies have higher $f_{\rm{q}}$, similar to the results of \citet{mcgee11};
\item There is no measurable difference in the quiescent fraction of galaxies in low ($10^{13} \lesssim M_{200} < 6 \times 10^{13} M_{\odot}$) and high ($6 \times 10^{13} M_{\odot} < M_{200} \lesssim 10^{14.5}$) mass groups at all stellar masses;
\item The fraction of groups with non-Gaussian velocity distributions increases from $z \sim 0$ to $z \sim 0.4$, while the fraction of groups with detected substructure is constant out to $z \sim 1$;
\item We observe the Butcher-Oemler effect in groups, where groups at higher redshifts have lower quiescent fractions, but only for low mass galaxies ($10 < \log_{10}(M_{\rm{star}}/M_{\odot}) \lesssim 10.5$), while galaxies with $\log_{10}(M_{\rm{star}}/M_{\odot}) > 10.5$ show little or no evidence of the Butcher-Oemler effect out to $z \sim 1$;
\item We do not observe a significant difference in the quiescent fractions of massive galaxies ($\log_{10}(M_{\rm{star}}/M_{\odot}) > 10.5$) in dynamically complex and relaxed groups, where the dynamical state is defined either by the AD or DS Test;
\item We observe a marginally lower quiescent fraction for low mass galaxies ($10 \leq \log_{10}(M_{\rm{star}}/M_{\odot}) < 10.5$) in groups with detected substructure at $z \sim 0.4$ when compared to groups with no significant substructure.
\end{enumerate}

In conclusion, we find that there is no strong correlation between the dynamical state of a group and the observed quiescent fraction for massive galaxies; however, we do see possible signs of a correlation between $f_{q}$ and substructure at $z \sim 0.4$.  This result suggests that environmental effects on galaxy evolution are only evident in low mass galaxies.  In order to better understand the role of group dynamics, and the environment in general, on the evolution of galaxies it is necessary to probe lower mass galaxies ($\log_{10}(M_{\rm{star}}/M_{\odot}) < 10.5$) where these mechanisms likely dominate.

\section{Acknowledgments}
We would like to thank the CNOC2 team for the use of unpublished redshifts.  A.H, L.C.P, and W.E.H would like to thank the National Science and Engineering Research Council of Canada (NSERC) for funding.

\bibliographystyle{apj}
\bibliography{ahou_group_dynevo_arxiv}

\end{document}